\documentstyle[prb,aps,tighten,preprint,eqsecnum]{revtex}
\begin{document}
\title{Quantum Disordered Systems with a Direction}
\author{K.B. Efetov}
\address{{\it Ruhr-Universitaet-Bochum, Universitaetsstrasse 150,
D-44780 Bochum, Germany \\and L.D. Landau Institute for Theoretical
Physics, Moscow, Russia}}
\date{\today{}}
\maketitle

\begin{abstract}
Models of disorder with a direction (constant imaginary vector-potential)
are considered. These non-Hermitian models can appear as a result of  
computation for models of statistical physics using transfer matrix technique  
or describe non-equilibrium processes. Eigenenergies of non-Hermitian 
Hamiltonians are not necessarily real and a joint probability density 
function of complex eigenvalues can characterize basic properties of 
the systems. This function is studied using the supersymmetry technique and
a supermatrix $\sigma$-model is derived. The $\sigma$-model differs from 
already known by a new term. The zero-dimensional version of the 
$\sigma$-model turns out to be the same as that obtained recently for 
ensembles of random weakly non-Hermitian or asymmetric real matrices. Using a 
new parametrization for the supermatrix $Q$ the density of complex eigenvalues 
is calculated in $0D$ for both the unitary and orthogonal ensembles. The 
function is drastically different in these two cases. It is everywhere smooth 
for the unitary ensemble but has a $\delta$-functional contribution for 
the orthogonal one. This anomalous part means that a finite portion of 
eigenvalues remains real at any degree of the non-Hermiticity. All details 
of the calculations are presented.
\end{abstract}

\section{Introduction}

Physics of disordered metals and semiconductors has been attracting a
considerable attention during several decades. Various interesting phenomena
were discovered experimentally and found a theoretical explanation. Rather
simple models of a particle moving in a random potential can be used to
describe such different effects as Anderson localization\cite{anderson},
mesoscopic fluctuations\cite{altshuler,lee}, Integer Quantum Hall Effect\cite
{laughlin}, and many others.

Although the phenomena can occur already at a weak disorder, a simple
perturbation theory in the disorder potential is not sufficient for their
quantitative description. A proper theory is often based on summing certain
classes of diagrams (cooperons and diffusons)\cite{gorkov,abrahams} but in
more complicated cases one has to use essentially non-perturbative methods
like the supersymmetry technique\cite{efetov0} based on mapping of the
disorder models onto a supermatrix $\sigma $-model (for a recent review see
Ref.\cite{efetov} and references therein). A disordered physical system can
include a magnetic field, magnetic and spin-orbit impurities, etc. However,
these additional interactions are included into the calculational schemes
without considerable difficulties.

By now, the diagrammatic expansions and the supersymmetry technique give a
possibility of getting explicit results for most of the disorder problems.
In addition, the supersymmetry method was applied for calculations with
random matrices\cite{ver}, which resulted in application of the method in
nuclear physics and quantum chaos where the random matrix theory (RMT) had
been the basic computational tool (for a review see, e.g. Refs.\cite
{gianonni,bohigas,casati,mehta,haake}). Recently, a supermatrix $\sigma $%
-model was derived for ballistic billiards averaging over either rare
impurities\cite{muz} or energy\cite{andreev}. So, the way of studying all
these interesting problems appears quite clear, although in some cases one
can encounter certain technical difficulties.

The systems mentioned above are described by quantum mechanical Hermitian
Hamiltonians. After averaging over disorder the systems involved are
invariant with respect to inversion of coordinates. Sometimes, in order to
describe the decay width of eigenstates, non-Hermitian Hamiltonians are
used. This approach is popular in study of quantum dots coupled to leads. Of
course, the Hamiltonian of the whole system of the dot with the leads is
Hermitian but it is often convenient to exclude the leads from the
consideration by integrating out degrees of freedom related to the leads. As
a result of such an integration one comes to an effective non-Hermitian
Hamiltonian of the dot containing imaginary energies\cite{zirnbauer}. This
type of the non-Hermiticity can be easily included into the scheme of the
supersymmetry technique as well as into diagrammatic expansions and many
results have been obtained explicitly\cite{efetov}.

In a recent publication\cite{hatano} Hatano and Nelson considered another
type of non-Hermitian Hamiltonians with a disorder, namely, Hamiltonians
with a constant ``imaginary vector potential''. In other words, the
Hamiltonians contain not only the second order derivative over space
coordinate but also the first order derivative with a real coefficient. The
model appears as a result of mapping of flux lines in a $\left( d+1\right) $%
-dimensional superconductor to the world lines of $d$-dimensional bosons.
Columnar defects produced experimentally by energetic heavy ion radiation%
\cite{civale} in order to pin the flux lines lead to the random potential in
the boson system, whereas the component of the magnetic field perpendicular
to the defects results in the constant imaginary vector potential\cite
{nelson}.

Already qualitative arguments\cite{hatano} indicate that the presence of the
imaginary vector potential can lead to new effects. In particular, a
one-dimensional chain of the bosons has to undergo a
localization-delocalization transition; this result was also checked by a
numerical computation. In ``conventional'' (without the first order
derivative) disordered systems transitions in one dimension do not occur and
therefore the model with a direction belongs to a really new class of
systems that have not been studied yet. It is argued that the localized
states should have real eigenenergies whereas eigenenergies of the extended
eigenstates may have a non-zero imaginary part.

The importance of investigation of such systems becomes even more evident if
one recalls that e.g. the equation for heat transfer with a convection has a
term with the first order derivative. One can imagine a situation when
quantum hopping of a particle from site to site of a lattice has a different
probability depending on direction. The presence of the first order
derivative in the Hamiltonian just corresponds to the introduction of a
certain direction. The non-equivalence of the directions can be provided by
coupling to another subsystem with broken inversion symmetry playing the
role of a reservoir; this reservoir may be out of equilibrium. The classical
analog of the disordered models with a direction (so called, directed
percolation) has been discussed in the literature\cite{obukhov}.

Another problem where one comes to a stochastic equation containing first
order derivatives is the problem of turbulence in flow dynamics. It is
generally believed that the most important features of the turbulence can be
described by the so called noisy Burgers equation\cite{burgers}, which is a
non-linear equation with a white noise random force. Besides its application
in the flow dynamics this equation is used as a toy model by field theorists
due to a striking analogy between the constant flux states in turbulence and
some anomalies in quantum field theories\cite{polyakov}. The Burgers
equation is equivalent to the Kardar-Parisi-Zhang equation introduced to
describe the crystal growth\cite{kardar}. The non-linear Burgers equation
can be reduced through a Hopf-Cole transformation to a linear $\left(
d+1\right) $ dimensional equation with a random potential and time playing
the role of the additional dimension. This equation has a first order time
derivative and there have already been an attempt to solve it using the
replica method\cite{bouchaud}. The noisy Burgers equation can also be
reduced to a quantum spin model with a non-Hermitian Hamiltonian\cite
{fogedby}. Recently, some interesting results have been obtained for the
Burgers equation using an ``instanton'' approximation\cite{gurarie}.

Independently of the study of the stochastic models with a direction a
considerable attention was paid in the last decade to investigation of
models of random real asymmetric and complex non-Hermitian matrices.
Eigenvalues of such matrices are generally speaking complex and so these
models are quite different from models of random real symmetric or Hermitian
matrices. Starting from the first work in this direction\cite{ginibre} a
number of publications\cite{girko,grobe,sommers,leh,haake} contain
discussion of properties of these models. Complex random matrices appear in
study of dissipative quantum maps\cite{grobe,haake} while real asymmetric
random matrices have found applications in neural network dynamics\cite
{som,doyon}. Many interesting aspects of non-Hermitian matrices were
discussed in preprints\cite{janik,feinberg}. Very recently a new regime of a
weak non-Hermiticity was found for complex random matrices\cite{fyodorov}.
In this regime an explicit formula for the density of complex eigenvalues
was obtained by mapping the problem onto a zero-dimensional supermatrix $%
\sigma $-model.

Although one may guess that the models with the non-Hermitian or real
asymmetric matrices should be related to disordered systems with
non-Hermitian Hamiltonians, no convincing arguments have been given as yet.
In fact, generally this is not true because, e.g. the models of open quantum
dots described by non-Hermitian Hamiltonians can hardly correspond to the
models of random non-Hermitian matrices discussed in the literature\cite
{ginibre,leh,haake,fyodorov}. However, as will be shown later, such a
correspondence does exist in some limiting cases for the disorder models
with a direction.

The goal of the present publication is to develop a method that would allow
to make analytical calculations for the disordered problems with a
direction. This goal is achieved by modifying the supersymmetry technique in
a way to include in the non-linear supermatrix $\sigma $-model terms
corresponding to the imaginary vector potential. Although a proper $\sigma $%
-model for the physical real vector potential has been derived long ago\cite
{efetov0}, changing to the imaginary one is far from trivial and, as a
result, a completely new term in the $\sigma $-model appears. The
zero-dimensional version of the $\sigma $-model turns out to be exactly the
same the one obtained in Ref.\cite{fyodorov} for the model of weakly
non-Hermitian random matrices.

The supermatrix $\sigma $-model derived below is valid in any dimension and
can be a proper tool for studying the localization-delocalization
transitions in one and two dimensions proposed in Ref.\cite{hatano}.
However, although one can use standard computational schemes\cite{efetov},
the presence of new terms in the $\sigma $-model make calculations with the
known parametrizations of the supermatrix $Q$ more difficult. Therefore, a
new parametrization is suggested and corresponding Jacobians are calculated.
To avoid ``overloading'' only zero-dimensional case is considered in this
article. For the unitary ensemble the result of Ref.\cite{fyodorov} for the
density of complex eigenvalues of weakly non-Hermitian random matrices is
reproduced. The density function is a smooth function of the imaginary part
of the eigenvalues, which shows that the probability of real eigenvalues is
zero.

In contrast, the density function for the orthogonal ensemble obtained below
contains a $\delta $-function, which shows that the fraction of states with
real eigenvalues is finite. This is a new very unusual and interesting
result. The entire function of the density of complex eigenvalues is
obtained for the first time. In the limit of strong non-Hermiticity the
probability functions for the both unitary and orthogonal ensembles
correspond to the ``elliptic law''\cite{ginibre,girko}.

The basic results of this article have been presented in a short form
elsewhere\cite{efetov1}. The article is organized as follows:

In Section II models of disorder with a direction are introduced and their
basic properties are discussed. Section III contains derivation of a
supermatrix $\sigma $-model. In Section IV a joint probability density of
complex eigenvalues is calculated for systems in a limited volume with
broken time reversal symmetry (unitary ensemble). This is done by
calculation of integrals over supermatrix $Q$ for the unitary ensemble. A
new parametrization for the supermatrices $Q$ is introduced. In Section V
similar calculations are carried out for the orthogonal ensemble. The result
for the density of complex eigenvalues proves to be qualitatively different
from that for the unitary ensemble. Section VI contains a discussion of the
results obtained and comparison with some other works. In Appendix the
Jacobians corresponding to the new parametrizations for the supermatrix $Q$
are derived.

\section{The model and its basic properties}

The initial classical model of vortices in a $\left( d+1\right) $%
-dimensional superconductor with line defects considered in Refs.\cite
{hatano,nelson} contains an interaction between the vortices. In the
corresponding quantum model of $d$-dimensional boson this describes an
interaction between the bosons. The interaction is, in principle, very
important. Its short range part does not allow bosons to condense at one
localized state. At the same time if it is strong enough there can be only
one boson in a localized state and the problem maps onto the model of
non-interacting fermions. Of course, this is not true for extended states
for which one should use the model of interacting bosons.

It is clear that one should first understand which one particle states are
localized and which are not. Therefore, as in Refs.\cite{hatano,nelson}, it
is reasonable to start with a $d$-dimensional Hamiltonian $H$ of
non-interacting particles including a constant imaginary vector potential $i%
{\bf h}$ and random potential of impurities $U\left( {\bf r}\right) $ 
\begin{equation}
\label{a1}H=H_0+U\left( {\bf r}\right) ,\;H_0=\frac{\left( {\bf \hat p+}i%
{\bf h}\right) ^2}{2m} 
\end{equation}
where ${\bf \hat p=-}i{\bf \nabla }$ and $m$ is the mass of a particle
(boson or fermion).

The random potential $U\left( {\bf r}\right) $ is assumed to be distributed
according to the Gaussian $\delta $-correlated law 
\begin{equation}
\label{b1}\left\langle U\left( {\bf r}\right) \right\rangle =0,\text{\ }%
\left\langle U\left( {\bf r}\right) U\left( {\bf r}^{\prime }\right)
\right\rangle =\frac 1{2\pi \nu \tau }\delta \left( {\bf r-r}^{\prime
}\right) 
\end{equation}
where $\tau $ is the mean free time, $\nu $ is the density of states at the
energy $\epsilon $ involved. As has been mentioned in the Introduction, the
potential $U\left( {\bf r}\right) $ corresponds to the potential of the line
defects and ${\bf h}$ to the component of the magnetic field for the model
of the vortices. At the same time, the Hamiltonian $H$, Eq. (\ref{a1}) can
describe other systems as well. So, we may study properties of the
Hamiltonian $H$ without recalling each time where it comes from. Some of
possible applications of Eq. (\ref{a1}) have been listed in the
Introduction. The directed quantum hopping appears a new interesting
possibility. The Hamiltonian $H_L$ of a lattice version of Eq. (\ref{a1})
can be written as follows 
\begin{equation}
\label{a2}H_L=-\frac t2\sum_{{\bf r}}\sum_{\nu =1}^d\left( e^{{\bf he}_\nu
}c_{{\bf r+e}_\nu }^{+}c_{{\bf r}}+e^{-{\bf he}_\nu }c_{{\bf r}}^{+}c_{{\bf %
r+e}_\nu }\right) +\sum_{{\bf r}}U\left( {\bf r}\right) c_{{\bf r}}^{+}c_{%
{\bf r}} 
\end{equation}
where $c^{+}$ and $c$ are creation and annihilation operators and $\left\{ 
{\bf e}_\nu \right\} $ are the unit lattice vectors.

Although Eq. (\ref{a2}) was used in Ref.\cite{hatano} only for numerical
calculations, it has a clear physical application. It describes quantum
hopping of a particle from site to site in the presence of a random
potential. However, the hopping probability along ${\bf h}$ is higher than
in the opposite direction. In other words, the Hamiltonian $H_L$ describes a
directed hopping in a random potential. The systems with the Hamiltonians $H$%
, $H_L$, Eqs. (\ref{a1},\ref{a2}), are not invariant with respect to
inversion of the coordinates even after averaging over impurities. At the
same time, they are time reversal invariant and therefore essentially
different from systems with real magnetic fields.

If necessary the Hamiltonians $H$ and $H_L$ can be generalized to include
the vector potential ${\bf A}$ corresponding to a physical magnetic field.
This can be done by the standard replacement 
\begin{equation}
\label{a3}{\bf \hat p\rightarrow \hat p-}\frac ec{\bf A} 
\end{equation}
in Eq. (\ref{a1}). Proper changes can also be done in Eq. (\ref{a2}).

Of course, the vortex model of Ref.\cite{hatano} corresponds to Eq. (\ref{a1}%
) with ${\bf A=}0$ but already the hopping model can be considered in an
arbitrary magnetic field. Changing the magnetic field (or, more precisely,
the vector potential ${\bf A}$) results in a crossover between ensembles
with different symmetries. In analogy with ``conventional'' (non-directed)
disordered systems these ensembles will be called orthogonal and unitary.

Although the Hamiltonians $H$ and $H_L$, Eqs. (\ref{a1}, \ref{a2}) are not
Hermitian this fact does not contradict to fundamental laws of nature. In
the problem of the vortices in superconductors these Hamiltonians appear
after a reduction of a $\left( d+1\right) $-dimensional classical problem to
a $d$-dimensional quantum one using the transfer matrix technique, which is
a formal trick. As concerns the directed hopping model the vector ${\bf h}$
can appear as a result of a coupling with another system (reservoir) which
is not necessarily in equilibrium. The latter system can be subjected e.g.
to an electric field, there can be non-decaying currents in it, etc.
Integrating out degrees of freedom related to the reservoir one obtains an
effective Hamiltonian that does not need to be Hermitian.

In other words, the non-Hermitian Hamiltonians appear at intermediate steps
of calculations and manipulations with them should be considered merely as
formal computational tricks. The corresponding wave functions and
eigenenergies are only formal objects as well. Of course, one should
understand how to relate initial physical observables to quantities
calculated with the non-Hermitian Hamiltonians.

It is relevant to mention that a classical directed model that can be
considered as the counterpart of the directed quantum problem has been
introduced long ago\cite{obukhov}. This is the model of a directed
percolation that can describe, e.g. spreading of infection or fire in a
forest affected by wind. According to a discussion of Ref.\cite{obukhov}
critical behavior near the percolation transition in the model of the
directed percolation is different from that of an isotropic model. The
analysis of Ref.\cite{obukhov} was based on a diagrammatic expansion. The
bare Green functions $G^{\left( 0\right) }\left( {\bf p}\right) $ used in
the expansion had the form 
\begin{equation}
\label{a4}G^{\left( 0\right) }\left( {\bf p}\right) =\frac 1{{\bf p}^2-i{\bf %
ap+}r} 
\end{equation}
with a constant vector ${\bf a}$. Comparing Eq.(\ref{a4}) with Eq. (\ref{a1}%
) we see that $G^{\left( 0\right) }$ is the Green function of the
Hamiltonian $H_0$, which demonstrates that both models are really closely
related to each other.

Now, let us discuss following Ref.\cite{hatano} basic properties of
eigenstates of the Hamiltonian $H$, Eq. (\ref{a1}). Due to the
non-Hermiticity of the Hamiltonian one should distinguish between right $%
\phi _k\left( {\bf r}\right) $ and left $\bar \phi _k\left( {\bf r}\right) $
eigenfunctions. They obey the following equations 
\begin{equation}
\label{a5}H\phi _k\left( {\bf r}\right) =\epsilon _k\phi _k\left( {\bf r}%
\right) ,\;H^T\bar \phi _k\left( {\bf r}\right) =\epsilon _k\bar \phi
_k\left( {\bf r}\right) 
\end{equation}
where $H^T$ is obtained by transposition of the Hamiltonian $H$. For
spinless particles the operation of the transposition means simply changing
of the sign of the space derivative. The functions $\bar \phi _k\left( {\bf r%
}\right) $ are also called conjugate to $\phi _k\left( {\bf r}\right) $; for
each eigenfunction one can construct its conjugate. The scalar product $%
\left( \phi _k,\phi _{k^{\prime }}\right) $ of two eigenfunctions $\phi
_k\left( {\bf r}\right) $ and $\phi _{k^{\prime }}\left( {\bf r}\right) $ is
introduced as 
\begin{equation}
\label{a6}\left( \bar \phi _k,\phi _{k^{\prime }}\right) =\int \bar \phi
_k\left( {\bf r}\right) \phi _{k^{\prime }}\left( {\bf r}\right) d{\bf r} 
\end{equation}
Using Eq. (\ref{a6}) one can prove in a standard way the orthogonality of
eigenfunctions corresponding to different eigenenergies. Together with the
normalization condition this can be written as 
\begin{equation}
\label{a7}\int \bar \phi _k\left( {\bf r}\right) \phi _{k^{\prime }}\left( 
{\bf r}\right) d{\bf r=}\delta _{kk^{\prime }} 
\end{equation}
The eigenenergy $\epsilon _k$ in both Eqs. (\ref{a5}) is the same. Eq. (\ref
{a7}) enables us to reproduce basic properties of conventional (Hermitian)
quantum mechanics replacing everywhere complex conjugates $\phi _\kappa
^{*}\left( {\bf r}\right) $ of the functions $\phi _k\left( {\bf r}\right) $
by the conjugates $\bar \phi _k\left( {\bf r}\right) $. However, the
eigenenergies $\epsilon _k$ in the non-Hermitian quantum mechanics are not
necessarily real. They must be real only if the functions $\phi _k^{*}\left( 
{\bf r}\right) $ and $\bar \phi _k^{}\left( {\bf r}\right) $ coincide. In
order to obtain well defined wave functions in the thermodynamic limit it is
convenient to impose periodic boundary conditions.

To understand better how the wave functions look like in different
situations it is instructive to consider a localized state with a
localization center at a point $x_0$ and extended states in the absence of
impurities (for simplicity we may restrict ourselves with the purely one
dimensional case). Assume that for $h=0$ the eigenfunctions $\phi _k^{\left(
0\right) }$ and the eigenvalues $\epsilon _k^{\left( 0\right) }$ are known.
Then, the functions 
\begin{equation}
\label{a8}\phi _k\left( x\right) =e^{hx}\phi _k^{\left( 0\right) }\left(
x\right) ,\;\bar \phi _k=e^{-hx}\phi _k^{\left( 0\right) }\left( x\right) 
\end{equation}
are solutions of Eqs. (\ref{a5}) with the eigenenergy $\epsilon _k^{\left(
0\right) }$.

At the same time, in order to satisfy the boundary conditions the function $%
\phi _k$ and $\bar \phi _k$ may not grow. If the function $\phi _k^{\left(
0\right) }\left( x\right) $ is exponentially localized at a distance $l_c$,
the function $\phi _k\left( x\right) $ takes the form 
\begin{equation}
\label{a9}\phi _k\left( x\right) =C\exp \left( h\left( x-x_0\right)
-l_c^{-1}\left| x-x_0\right| \right) 
\end{equation}
The function $\phi _k\left( x\right) $, Eq. (\ref{a9}), and the
corresponding function $\bar \phi _k\left( x\right) $ does not grow at $%
\left| x\right| \rightarrow \infty $ only if $\left| h\right| <l_c^{-1}$.
The point $\left| h\right| =l_c^{-1}$ was identified\cite{hatano} with a
localization-delocalization transition.

In the region $\left| h\right| \geq l_c^{-1}$ the functions $\phi _k$ given
by Eqs. (\ref{a8}, \ref{a9}) are longer eigenfunctions because they do not
satisfy the boundary conditions. To get an idea how the eigenfunctions look
like in this region we may neglect the disorder potential. Then, the plane
waves 
\begin{equation}
\label{a10}\phi _k=L^{-1/2}e^{ikx},\;\bar \phi _k=L^{-1/2}e^{-ikx} 
\end{equation}
where $L$ is the length of the sample, are proper solutions of Eqs. (\ref{a5}%
) satisfying the boundary conditions. However, in this case the eigenvalue $%
\epsilon _k$ is no longer real 
\begin{equation}
\label{a11}\epsilon _k=\frac{\left( k+ih\right) ^2}{2m} 
\end{equation}
We see that the question about whether an eigenfunction in the presence of
the imaginary vector potential is localized or extended is closely related
in the thermodynamic limit to the question whether the corresponding
eigenenergy is real or complex. The arguments presented are qualitative but
they were confirmed by numerical calculations\cite{hatano}.

It is clear from the previous discussion that it is very important to
understand when eigenenergies are real and when they become complex. A
convenient function characterizing the system is the joint probability
density of complex eigenenergies $P\left( \epsilon ,y\right) $ defined as 
\begin{equation}
\label{a12}P\left( \epsilon ,y\right) =\frac 1V\left\langle \sum_k\delta
\left( \epsilon -\epsilon _k^{\prime }\right) \delta \left( y-\epsilon
_k^{\prime \prime }\right) \right\rangle 
\end{equation}
where $\epsilon _k^{\prime }$ and $\epsilon _k^{\prime \prime }$ are the
real and imaginary parts of the eigenenergy $\epsilon _k$, $V$ is the volume
and the angle brackets stand for averaging over impurities. If all states
are localized, such that $\epsilon _k^{\prime \prime }=0$, the function $%
P\left( \epsilon ,y\right) $ equals 
\begin{equation}
\label{a13}P\left( \epsilon ,y\right) =\nu \left( \epsilon \right) \delta
\left( y\right) 
\end{equation}
where $\nu \left( \epsilon \right) $ is the average density of states.

If all states are extended the function $P\left( \epsilon ,y\right) $ should
be a smooth function of both variables. In some cases physical quantities
can be expressed directly through the function $P\left( \epsilon ,y\right) $
although other correlation functions are also of interest. The rest of this
article is devoted to reduction of the function $P\left( \epsilon ,y\right) $%
, which is the simplest non-trivial function characterizing the system, to a
correlation function in a supersymmetric $\sigma $-model and to some
calculations with this model. This is the first attempt of a quantitative
analytical study of the disordered directed quantum systems.

\section{Derivation of $\sigma $-model}

According to the standard procedure of derivation of the supermatrix $\sigma 
$-model\cite{efetov0,efetov} one should express the physical quantity in
terms of retarded $G_\epsilon ^R$ and advanced $G_\epsilon ^A$ Green
functions of the Hamiltonian. Usually the average density of states that can
be expressed through the average of one Green function is not an interesting
quantity because it does not distinguish between localized and extended
states. The density of complex eigenvalues $P\left( \epsilon ,y\right) $ is
definitely more interesting but how to express it in terms of integrals over
supervectors, which is the first step of derivation of the $\sigma $-model?

The problem is that it is not clear how to write the function $P\left(
\epsilon ,y\right) $ in terms of the functions $G_\epsilon ^R$, $G_\epsilon
^A$. However, even if this representation existed it would not help. Using
the spectral expansion of the functions $G^{R,\grave A}$ 
\begin{equation}
\label{a14}G_\epsilon ^{R,A}\left( {\bf r,r}^{\prime }\right) =\sum_k\frac{%
\phi _k\left( {\bf r}\right) \bar \phi _k\left( {\bf r}^{\prime }\right) }{%
\epsilon -\epsilon _k\pm i\delta } 
\end{equation}
we see that if some eigenenergies $\epsilon _k$ are complex the function $%
G_\epsilon ^R\left( G_\epsilon ^A\right) $ is no longer analytical in the
upper (lower) half plane of complex $\epsilon $. But the very possibility to
rewrite the Green functions in terms of convergent Gaussian integrals over
the supervectors was based on the assumption that the eigenenergies were
real.

Another possibility is based on the relation 
\begin{equation}
\label{a15}\delta \left( a\right) \delta \left( b\right) =\frac 1\pi \lim
_{\gamma \rightarrow 0}\frac{\gamma ^2}{\left( a^2+b^2+\gamma ^2\right) ^2} 
\end{equation}
that holds for real $a$ and $b$. With Eq. (\ref{a15}) the density function $%
P\left( \epsilon ,y\right) $ can be rewritten as 
\begin{equation}
\label{a16}P\left( \epsilon ,y\right) \text{=}\frac{\gamma ^2}{\pi V}\lim
_{\gamma \rightarrow 0}\left\langle \sum_k\left[ \left( \epsilon -\epsilon
_k^{\prime }\right) ^2+\left( y-\epsilon _k^{\prime \prime }\right)
^2+\gamma ^2\right] ^{-2}\right\rangle 
\end{equation}
Using the orthogonality of the eigenfunctions $\phi _k$ Eq. (\ref{a16}) can
be also represented as 
\begin{equation}
\label{a17}P\left( \epsilon ,y\right) =\frac 1{\pi V}\lim _{\gamma
\rightarrow 0}\int B\left( {\bf r,r}^{\prime }\right) B\left( {\bf r}%
^{\prime },{\bf r}\right) d{\bf r}d{\bf r}^{\prime } 
\end{equation}
where the function $B\left( {\bf r,r}^{\prime }\right) $ has the form 
\begin{equation}
\label{a18}B\left( {\bf r,r}^{\prime }\right) =\sum_k\frac{\gamma \phi
_k\left( {\bf r}\right) \bar \phi _k\left( {\bf r}^{\prime }\right) }{\left(
\epsilon -\epsilon _k^{\prime }\right) ^2+\left( y-\epsilon _k^{\prime
\prime }\right) ^2+\gamma ^2}\text{ } 
\end{equation}
The representation of the density function $P\left( \epsilon ,y\right) $ in
by Eq. (\ref{a17}) is very convenient because it allows to rewrite this
function in terms of a Gaussian integral over supervectors.

In order to derive a proper expression let us introduce an Hermitian
operator $\hat M$ 
\begin{equation}
\label{a19}\hat M=\left( 
\begin{array}{cc}
H^{\prime }-\epsilon & i\left( H^{\prime \prime }-y\right) \\ 
-i\left( H^{\prime \prime }-y\right) & -\left( H^{\prime }-\epsilon \right) 
\end{array}
\right) 
\end{equation}
where 
\begin{equation}
\label{a20}H^{\prime }=\frac 12\left( H+H^{+}\right) ,\qquad H^{\prime
\prime }=-\frac i2\left( H-H^{+}\right) 
\end{equation}
In Eq. (\ref{a20}) the symbol ``$+$'' means Hermitian conjugation. For real
Hamiltonians this conjugation coincides with the transposition ``$T$''.
However, let us write formulae in a general form such that the Hamiltonian $%
H $ may include magnetic interactions and be complex.

Instead of manipulating with the non-Hermitian operator $H$ one can try to
use the Hermitian operator $\hat M$. To follow the standard procedure of the
supersymmetry technique one should find first the eigenstates of this
operator. For the complex non-Hermitian operator $H$ one can write $4$
equations for the eigenstates 
\begin{equation}
\label{a21}H\phi _k=\epsilon _k\phi _k,\qquad H^T\bar \phi _k=\epsilon _k%
\bar \phi _k 
\end{equation}

\begin{equation}
\label{a22}H^{*}\phi _k^{*}=\epsilon _k^{*}\phi _k^{*},\qquad H^{+}\bar \phi
_k^{*}=\epsilon _k^{*}\bar \phi _k^{*} 
\end{equation}
Eqs. (\ref{a22}) are merely complex conjugates of Eqs. (\ref{a21}).

Now, let us introduce two sets of $2$-component vectors $u_k$ and $v_k$ 
\begin{equation}
\label{a23}u_k=\frac 12\left( 
\begin{array}{c}
\phi _k+ 
\bar \phi _k^{*} \\ \phi _k-\bar \phi _k^{*} 
\end{array}
\right) ,\qquad v_k=\frac 12\left( 
\begin{array}{c}
\phi _k- 
\bar \phi _k^{*} \\ \phi _k+\bar \phi _k^{*} 
\end{array}
\right) , 
\end{equation}
$$
\bar u_k=\frac 12\left( 
\begin{array}{cc}
\bar \phi _k+\phi _k^{*} & \bar \phi _k-\phi _k^{*} 
\end{array}
\right) ,\quad \bar v_k=\frac 12\left( 
\begin{array}{cc}
\bar \phi _k-\phi _k^{*} & \bar \phi _k+\phi _k^{*} 
\end{array}
\right) 
$$
Using the orthogonality of the eigenfunctions $\phi _k$, Eq. (\ref{a7}), one
can prove the orthogonality of the vectors $u_k$ and $v_k$%
\begin{equation}
\label{a24}\int \bar u_k\left( {\bf r}\right) u_{k^{\prime }}\left( {\bf r}%
\right) d{\bf r=}\int \bar v_k\left( {\bf r}\right) v_{k^{\prime }}\left( 
{\bf r}\right) d{\bf r=}\delta _{kk^{\prime }} 
\end{equation}

$$
\int \bar u_k\left( {\bf r}\right) v_{k^{\prime }}\left( {\bf r}\right) d%
{\bf r=}\int \bar v_k\left( {\bf r}\right) u_{k^{\prime }}\left( {\bf r}%
\right) d{\bf r=}0 
$$
It is not difficult to see that the vectors $u_k\left( {\bf r}\right) $ and $%
v_k\left( {\bf r}\right) $ are eigenvectors of the matrix operator $\hat M$
satisfying the equations 
\begin{equation}
\label{a25}\hat Mu_k=M_ku_k,\qquad \hat Mv_k=M_kv_k 
\end{equation}
where the matrix $M_k$ equals 
\begin{equation}
\label{a26}M_k=\left( 
\begin{array}{cc}
\epsilon _k^{\prime }-\epsilon & i\left( \epsilon _k^{\prime \prime
}-y\right) \\ 
-i\left( \epsilon _k^{\prime \prime }-y\right) & -\left( \epsilon _k^{\prime
}-\epsilon \right) 
\end{array}
\right) 
\end{equation}
and $\epsilon _k^{\prime }$, $\epsilon _k^{\prime \prime }$ are the real and
imaginary parts of the eigenenergies $\epsilon _k$.

Using the identity 
\begin{equation}
\label{a27}\frac i2Tr\left( M_k+i\gamma \right) ^{-1}=\frac \gamma {\left(
\epsilon _k^{\prime }-\epsilon \right) ^2+\left( \epsilon _k^{\prime \prime
}-y\right) ^2+\gamma ^2} 
\end{equation}
one can see that the functions $B\left( {\bf r,r}^{\prime }\right) $, Eq. (%
\ref{a18}), are closely related to the operator $\hat M$. The only thing
that remains to be done is to express the matrix $\left( M_k+i\gamma \right)
^{-1}$ and then the operator $\left( \hat M+i\gamma \right) ^{-1}$ in terms
of a Gaussian integral over supervectors.

The operator $\hat M$ is Hermititian, its eigenvectors $u_k$ and $v_k$, Eqs.
(\ref{a23}) are known and therefore we can follow the standard procedure of
the derivation\cite{efetov0,efetov}. Changing from the Hamiltonian $H$ to
the operator $\hat M$ we had to double the size of the relevant matrices.
This means that in order to write proper Gaussian integrals we should use,
as usually, $8$-component supervectors $\psi \left( {\bf r}\right) $. In
fact, one comes to supervectors $\psi $ with exactly the same structure as
previously\cite{efetov0,efetov} 
\begin{equation}
\label{a28}\psi ^m=\left( 
\begin{array}{c}
\vartheta ^m \\ 
r^m 
\end{array}
\right) ,\quad \vartheta ^m=\frac 1{\sqrt{2}}\left( 
\begin{array}{c}
\chi ^{m*} \\ 
\chi ^m 
\end{array}
\right) ,\quad r^m=\frac 1{\sqrt{2}}\left( 
\begin{array}{c}
S^{m*} \\ 
S^m 
\end{array}
\right) 
\end{equation}
$m=1$,$2$; $\chi ^m$ and $S^m$ are anticommuting and commuting variables
respectively.

Let us present several important intermediate steps of the reduction of the
operator $\left( \hat M+i\gamma \right) ^{-1}$, to the functional integral
over $\psi \left( {\bf r}\right) $. First, we have 
\begin{equation}
\label{a29}\left( i\gamma +M_k\right) ^{-1}=-i\int \left(
a_ka_k^{*}+b_kb_k^{*}\right) \exp \left( -L_k\right) dR_k=-i\int \left(
\sigma _k\sigma _k^{*}+\rho _k\rho _k^{*}\right) \exp \left( -L_k\right)
dR_k 
\end{equation}
where $a_k,b_k$ and $\sigma _k,\rho _k$ are commuting and anticommuting
variables, respectively, $dR_k$ stands for the elementary volume in the
space of these variables. The function $L_k$ in Eq. (\ref{a29}) equals 
\begin{equation}
\label{a30}L_k=-i\left( 
\begin{array}{cc}
a_k^{*} & b_k^{*} 
\end{array}
\right) \left( i\gamma +M_k\right) \left( 
\begin{array}{c}
a_k \\ 
b_k 
\end{array}
\right) -i\left( 
\begin{array}{cc}
\sigma _k^{*} & \rho _k^{*} 
\end{array}
\right) \left( i\gamma +M_k\right) \left( 
\begin{array}{c}
\sigma _k \\ 
\sigma _k^{*} 
\end{array}
\right) 
\end{equation}
The vector fields $\vec \chi \left( {\bf r}\right) $ and $\vec S\left( {\bf r%
}\right) $ are introduced as 
\begin{equation}
\label{a31}\vec \chi \left( {\bf r}\right) =\left( 
\begin{array}{c}
\chi ^1\left( 
{\bf r}\right) \\ \chi ^2\left( {\bf r}\right) 
\end{array}
\right) =\sum_k\left( a_ku_k\left( {\bf r}\right) +b_kv_k\left( {\bf r}%
\right) \right) , 
\end{equation}

$$
\vec S\left( {\bf r}\right) =\left( 
\begin{array}{c}
S^1\left( 
{\bf r}\right) \\ S^2\left( {\bf r}\right) 
\end{array}
\right) =\sum_k\left( \sigma _ku_k\left( {\bf r}\right) +\rho _kv_k\left( 
{\bf r}\right) \right) 
$$
where the vectors $u_k$ and $v_k$ are defined in Eqs. (\ref{a23}).

With these definitions one can express the functions $B\left( {\bf r,r}%
^{\prime }\right) $, Eq. (\ref{a18}), in terms of Gaussian integrals over
the vector fields $\vec \chi \left( {\bf r}\right) $ and $\vec S\left( {\bf r%
}\right) $. The derivation is based on the identity 
\begin{equation}
\label{a32}\int \vec S^{*}\left( {\bf r}\right) \hat M\vec S\left( {\bf r}%
\right) d{\bf r=}\sum_k\left( 
\begin{array}{cc}
a_k^{*} & b_k^{*} 
\end{array}
\right) \hat M_k\left( 
\begin{array}{c}
a_k \\ 
b_k 
\end{array}
\right) 
\end{equation}
that can be proven using Eqs. (\ref{a7},\ref{a23}) (the same for $\vec \chi
\left( {\bf r}\right) $). Less trivial is the expression 
\begin{equation}
\label{a33}i\gamma \int \vec S^{*}\left( {\bf r}\right) \vec S\left( {\bf r}%
\right) d{\bf r} 
\end{equation}
Using the expansion, Eq. (\ref{a31}), we can see that the integral, Eq. (\ref
{a33}), contains non-diagonal with respect to $k,k^{\prime }$ terms. For
example, there is the following term 
\begin{equation}
\label{a34}\frac{i\gamma }2\sum_{k,k^{\prime }}a_{k^{\prime }}^{*}a_k\int
\phi _{k^{\prime }}^{*}\left( {\bf r}\right) \phi _k\left( {\bf r}\right) d%
{\bf r} 
\end{equation}
For Hermitian Hamiltonians the integral in Eq. (\ref{a34}) would give $%
\delta _{kk^{\prime }}$. However, generally it is not zero for arbitrary $k$
and $k^{\prime }$ because the orthogonality relation, Eq. (\ref{a7}),
contains $\bar \phi _k$ but not $\phi _k^{*}$. Fortunately, this does not
create difficulties in the limit of small ``vector potential'' $h$ that is
of the main interest in the present work, because the difference between $%
\bar \phi _k$ and $\phi _k^{*}$ is small. This allows us to write 
\begin{equation}
\label{a35}\sum_kL_k=\int [\vec \chi ^{*}\left( {\bf r}\right) \left(
i\gamma +\hat M\right) \vec \chi \left( {\bf r}\right) +\vec S^{*}\left( 
{\bf r}\right) \left( i\gamma +\hat M\right) \vec S\left( {\bf r}\right) ]d%
{\bf r} 
\end{equation}
Although one can use Eq. (\ref{a35}) as an effective Lagrangian, it is
convenient\cite{efetov0,efetov} to unify all components of the vectors $\vec 
\chi $, $\vec \chi ^{*}$, $\vec S$, and $\vec S^{*}$ into the supervector $%
\psi $ of the form, Eq. (\ref{a28}). As a result, one comes to integration
with the weight $\exp \left( -{\cal L}\right) $, where the Lagrangian ${\cal %
L}$ takes the form 
\begin{equation}
\label{a36}{\cal L}=-i\int \bar \psi \left( {\bf r}\right) \left( {\cal H}%
_0+U\left( {\bf r}\right) \right) \psi \left( {\bf r}\right) d{\bf r} 
\end{equation}
where the ``charge-conjugate'' supervector $\bar \psi \left( {\bf r}\right) $
is the same as in Refs.\cite{efetov0,efetov}. The $8\times 8$ matrix
operator ${\cal H}_0$ can be written as 
\begin{equation}
\label{a37}{\cal H}_0={\cal H}_{00}+{\cal H}_{01}, 
\end{equation}
$$
{\cal H}_{00}=H_0^{\prime }-\epsilon +i\gamma \Lambda ,\qquad {\cal H}%
_{01}=i\Lambda _1\left( H^{\prime \prime }+y\tau _3\right) 
$$
In the continuum model, the ``imaginary'' part $H^{\prime \prime }$ of the
Hamiltonian $H$, Eqs. (\ref{a1},\ref{a20}), has the form 
\begin{equation}
\label{a38}H^{\prime \prime }=-i\frac{{\bf h\nabla }}m 
\end{equation}
The diagonal matrices $\Lambda $ and $\tau _3$ are the same as in Refs.\cite
{efetov0,efetov}. The matrix $\Lambda _1$ anticommutes with the matrix $%
\Lambda $ and also consists of unit $4\times 4$ blocks. The explicit form of
these matrices is 
\begin{equation}
\label{a39}\Lambda =\left( 
\begin{array}{cc}
{\bf 1} & 0 \\ 
0 & -{\bf 1} 
\end{array}
\right) ,\qquad \Lambda _1=\left( 
\begin{array}{cc}
0 & {\bf 1} \\ {\bf 1} & 0 
\end{array}
\right) 
\end{equation}
Eq. (\ref{a36}) is similar to the corresponding equation for localization
problems\cite{efetov0,efetov} and in the absence of ${\cal H}_{01}$ these
equations would coincide. All new physics comes from the operator ${\cal H}%
_{01}$. A magnetic field can be included into ${\cal H}_{00}$ in a standard
way.

All subsequent manipulations are the same as in Refs.\cite{efetov0,efetov}.
First, one averages over the random potential $U\left( {\bf r}\right) $
using Eq. (\ref{b1}) and comes instead of Eq. (\ref{a36}) to a regular
Lagrangian ${\cal L}$%
\begin{equation}
\label{a40}{\cal L}=\int \left[ -i\bar \psi \left( {\bf r}\right) {\cal H}%
_0\psi \left( {\bf r}\right) +\frac 1{4\pi \nu \tau }\left( \bar \psi \left( 
{\bf r}\right) \psi \left( {\bf r}\right) \right) ^2\right] d{\bf r} 
\end{equation}
Then, one decouples the interaction term in Eq. (\ref{a40}) by integration
over a supermatrix $Q$ and integrates over the supervector $\psi $ assuming
that the supermatrix $Q$ varies in space slowly. After that one comes to an
integral over $Q$ with the weight $\exp \left( -F\left[ Q\right] \right) $.
The functional integral over $Q$ is calculated using a saddle-point
approximation. At the saddle-point the supermatrix $Q$ does not depend on
coordinates and in the limit of small ${\cal H}_{01}$ and $\gamma $ one
obtains the standard equation 
\begin{equation}
\label{a41}Q\left( {\bf r}\right) =\frac 1{\pi \nu }\left( \left[ -i{\cal H}%
_{00}+\frac{Q\left( {\bf r}\right) }{2\tau }\right] ^{-1}\right) _{{\bf rr}} 
\end{equation}
which leads to the constraint $Q^2=1$. Now, one has to expand the free
energy functional $F\left[ Q\right] $ near the saddle-point in {\em H}$_{01}$%
, $\gamma $ and $\nabla Q$. As a result the functional $F\left[ Q\right] $
acquires the form of a $\sigma $-model 
\begin{equation}
\label{a42}F\left[ Q\right] =\frac{\pi \nu }8\int STr[D_0\left( {\bf \nabla }%
Q+{\bf h}\left[ Q,\Lambda _1\right] \right) ^2-4\left( \gamma \Lambda
+y\Lambda _1\tau _3\right) Q]d{\bf r} 
\end{equation}
where $D_0$ is the classical diffusion coefficient, $\left[ .,.\right] $ is
commutator and $STr$ stands for supertrace. Eq. (\ref{a42}) is written in
the absence of a magnetic field. The expansion near the saddle-point leading
to Eq. (\ref{a42}) is justified provided $y\ll \tau ^{-1}$ and $h\ll l^{-1}$%
, where $l$ is the mean free path. The supermatrices $Q$ are the same as
those for the orthogonal ensemble\cite{efetov0,efetov}. This case can
correspond to the problem of vortices in superconductors with line defects%
\cite{hatano}. If for some other problems one has to include in the
Hamiltonians $H$, $H_L$ the physical vector-potential ${\bf A}$
corresponding to a magnetic field, the standard derivation shows that the
proper $\sigma $-model is obtained from Eq. (\ref{a42}) by the replacement 
\begin{equation}
\label{a43}{\bf \nabla }Q{\bf \rightarrow \nabla }Q-\frac{ie}c{\bf A}\left[
Q,\tau _3\right] 
\end{equation}
In the limit of a strong magnetic field one can neglect fluctuations of a
certain symmetry (cooperons). Then, Eq. (\ref{a42}) is still valid but the
supermatrices $Q$ should have the symmetry corresponding to the unitary
ensemble.

The free energy functional $F\left[ Q\right] $, Eq. (\ref{a42}) has two
additional with respect to the functional used for ``conventional'' disorder
problems. These terms contain the matrix $\Lambda _1$, which leads to new
effective ``external fields'' in the free energy. We see from Eqs. (\ref{a42}%
,\ref{a43}) that ${\bf h}$ and ${\bf A}$ enter $F\left[ Q\right] $ in a
different way. A simple replacement ${\bf A\rightarrow }i{\bf h}$ in the $%
\sigma $-model of Refs. (\cite{efetov0,efetov}) would give a wrong result.
This reflects the fact that a non-zero ${\bf A}$ violates the time reversal
symmetry while ${\bf h}$ can break only the symmetry with respect to
inversion of coordinates.

In order to write express the density function $P\left( \epsilon ,y\right) $%
, Eq. (\ref{a17},\ref{a18}), in terms of a functional integral over $Q$ one
should know not only the weight $\exp \left( -F\left[ Q\right] \right) $ but
also a pre-exponential functional $A\left[ Q\right] $. It can be derived
from Eqs. (\ref{a17},\ref{a18}) in a standard way. One of the functions $B$
can be written using the first line of Eq. (\ref{a29}) and the other using
the second one. As a result, one obtains in the pre-exponential a product of
four different components of the supervector $\psi $; two of them are at the
point ${\bf r}$ while the other two at the point ${\bf r}^{\prime }$. After
averaging over the random potential $U\left( {\bf r}\right) $ and decoupling
of the effective interaction in Eq. (\ref{a40}) by integration over the
supermatrix $Q$ one has to compute Gaussian integrals over $\Psi $. This can
be done using the Wick theorem. In the limit $\tau ^{-1}\ll \left( \nu
V\right) ^{-1}$ one may take into account only pairing of two $\psi $ at
coinciding points. The rest of the calculation is simple and one obtains 
\begin{equation}
\label{a44}P\left( \epsilon ,y\right) =-\lim _{\gamma \rightarrow 0}\frac{%
\pi \nu ^2}{4V}\int A\left[ Q\right] \exp \left( -F\left[ Q\right] \right)
dQ, 
\end{equation}
where 
\begin{equation}
\label{a45}A\left[ Q\right] =\int [\left( Q_{42}^{11}\left( {\bf r}\right)
+Q_{42}^{22}\left( {\bf r}\right) \right) \left( Q_{24}^{11}\left( {\bf r}%
^{\prime }\right) +Q_{24}^{22}\left( {\bf r}^{\prime }\right) \right) 
\end{equation}
$$
-\left( Q_{42}^{21}\left( {\bf r}\right) +Q_{42}^{12}\left( {\bf r}\right)
\right) \left( Q_{24}^{21}\left( {\bf r}^{\prime }\right) +Q_{24}^{12}\left( 
{\bf r}^{\prime }\right) \right) ]d{\bf r}d{\bf r}^{\prime } 
$$
Numeration of the matrix elements in Eq. (\ref{a45}) is standard\cite
{efetov0,efetov}.

Eqs. (\ref{a42}-\ref{a45}) solve the problem of mapping of the density of
complex eigenvalues for disorder models with a direction onto a supermatrix $%
\sigma $-model. The density function $P\left( \epsilon ,y\right) $ depends
on the real part $\epsilon $ of the eigenenergies through the parameters $%
\nu $ and $D_0$ that are dependent on $\epsilon $. The dependence on the
imaginary part $y$ is more complicated. Remarkably, the $\sigma $-model
derived differs from the $\sigma $-model for localization problems by
additional ``external fields'' only. This simplifies calculations because
one can use well developed computational schemes.

The $\sigma $-model, Eqs. (\ref{a42}-\ref{a45}) can be used in any
dimension. The one-dimensional version describes ``quantum wires'' or, in
the language of the superconductor model, to vortices in a slab. According
to a discussion of Ref.\cite{hatano}, in one-dimensional models there has to
be a localization-delocalization transition. If this is true for thick wires
the one-dimensional $\sigma $-model should undergo a phase transition when
changing the value of $h$. However, study of the one-dimensional model is
more difficult than of the zero-dimensional one. Leaving higher dimensional
problems for future investigation let us concentrate in the next Section on
calculating the density function $P\left( \epsilon ,y\right) $ for a sample
with a finite volume. This situation is described by the zero-dimensional $%
\sigma $-model.

\section{Density of complex eigenvalues in a limited volume: unitary ensemble
}

If disorder is not very strong there is a regime when physical quantities
can be obtained from the zero-dimensional ($0D$) $\sigma $-model. This is
the limiting case when one considers only supermatrices $Q$ that do not vary
in space. For the problem of level statistics in Hermitian models the $0D$ $%
\sigma $-model is obtained in the limit $\omega \ll E_c$, where $E_c=\pi
^2D_0/L_{}^2$ is the Thouless energy ($L$ is the sample size)\cite
{efetov0,efetov}. If the sample is connected with leads and the energy
levels are smeared the $0D$ case is possible provided the level width does
not exceed $E_c$. If the disorder is strong or the sample has one- or
two-dimensional geometry, such that the localization length $L_c$ is smaller
than the sample size, the $0D$ limit cannot be achieved.

It is clear that the situation with the directed problems involved should be
similar and one can come to the $0D$ $\sigma $-model provided $h$, $y$, and $%
\gamma $ in Eq. (\ref{a42}) are not very large and disorder is not very
strong. For the model of vortices in a superconductor the $0D$ limit for the 
$\sigma $-model would correspond to a sample with a finite cross-section
perpendicular to the line defects.

Neglecting all non-zero space harmonics in the free energy functional $%
F\left[ Q\right] $ one can rewrite Eq. (\ref{a42}) as follows 
\begin{equation}
\label{e1}F\left[ Q\right] =STr\left( \frac{a^2}{16}\left[ Q,\Lambda
_1\right] ^2-\frac x4\Lambda _1\tau _3Q-\frac{\tilde \gamma }4\Lambda
Q\right) 
\end{equation}
where 
\begin{equation}
\label{e2}a^2=\frac{2\pi D_0h^2}\Delta ,\qquad x=\frac{2\pi y}\Delta ,\qquad 
\tilde \gamma =\frac{2\pi \gamma }\Delta 
\end{equation}
and $\Delta $=$\left( \nu V\right) ^{-1}$ is the mean level spacing.

The distribution function $P\left( \epsilon ,y\right) $, Eqs. (\ref{a44},\ref
{a45}), takes the form 
\begin{equation}
\label{e3}P\left( \epsilon ,y\right) =-\frac{\pi \nu }{4\Delta }\lim _{%
\tilde \gamma \rightarrow 0}\int A\left[ Q\right] \exp \left( -F\left[
Q\right] \right) dQ, 
\end{equation}

$$
A\left[ Q\right] =\left( Q_{42}^{11}+Q_{42}^{22}\right) \left(
Q_{24}^{11}+Q_{24}^{22}\right) -\left( Q_{42}^{21}+Q_{42}^{12}\right) \left(
Q_{24}^{21}+Q_{24}^{12}\right) 
$$
with $F\left[ Q\right] $ determined by Eq. (\ref{e1}).

The non-zero space harmonics can be neglected provided the following
inequalities are fulfilled 
\begin{equation}
\label{e4}y\ll E_c,\quad \tilde \gamma \ll E_c,\quad h\ll L^{-1} 
\end{equation}
where $L$ is the sample size.

To obtain the function $P\left( \epsilon ,y\right) $ one should calculate in
Eq. (\ref{e3}) a definite integral over the supermatrices $Q$. The structure
of supermatrices $Q$ is the same as in Refs.\cite{efetov0,efetov} and, in
principle, the way how to compute the integral is clear. As usual, all
manipulations are simpler for the unitary ensemble and therefore let us
start with this case.

However, already before an explicit calculation of the integral in Eq. (\ref
{e3}) an interesting observation can be made. We know that the $0D$ version
of the $\sigma $-model for Hermitian disordered systems can also be derived
from random matrix models\cite{ver}. In fact, it is the way how the
equivalence of between disordered systems in a limited volume and random
matrix theory (RMT) was finally established. Now, a natural question arises:
do the random models with a direction considered in the present work
correspond to a RMT?

Of course, this cannot be a model of Hermitian or real symmetric matrices
because in this case all eigenvalues must be real. So, one should think of
ensembles of random real asymmetric or complex non-Hermitian matrices. Study
of random complex matrices without the requirement of Hermiticity has
started quite long ago\cite{ginibre} and since then models of non-Hermitian
or real asymmetric random matrices have been considered in a number of
publications\cite{mehta,haake,girko,grobe,leh,som,doyon}. The ensembles of
real symmetric random matrices have found applications in e.g. neural
network dynamics\cite{som,doyon} while the ensembles of complex random
matrices appear in study of dissipative quantum maps\cite{grobe,haake}. One
of results obtained is that, for Gaussian ensembles in the limit of a large
size $N$ of the matrices, the eigenvalues are uniformly distributed in an
ellipse\cite{girko,leh,sommers}.

Recently, an ensemble of ``weakly non-Hermitian'' random matrices $X$ was
introduced\cite{fyodorov}. It was assumed that these matrices had the form 
\begin{equation}
\label{e5}\hat X=\hat A+i\alpha N^{-1/2}\hat B 
\end{equation}
with $N\times N$ statistically independent Hermitian matrices $A$ and $B$,
and a number $\alpha $ of the order of unity. The matrices $\hat A$ and $%
\hat B$ obeyed Gaussian distributions with the probability densities 
\begin{equation}
\label{e6}{\cal P}\left( \hat A\right) \propto \exp \left( -\frac N{2J^2}Tr%
\hat A^2\right) ,\qquad {\cal P}\left( \hat B\right) \propto \exp \left( -%
\frac N{2J^2}Tr\hat B^2\right) 
\end{equation}
where $J$ has order of unity.

The parameter $\alpha N^{-1/2}$ is a measure of the non-Hermiticity and is
always small for $N\rightarrow \infty $ and $\alpha $ finite. The authors of
Ref.\cite{fyodorov} calculated a density of complex eigenvalues similar to
the function $P\left( \epsilon ,y\right) $, Eq. (\ref{a12}) and demonstrated
that this function has a finite limit when $N\rightarrow \infty $. At the
same time they did not point out any direct physical applications. For
computation of the function $P\left( \epsilon ,y\right) $ they used the
supersymmetry technique. Remarkably, a $\sigma $-model derived in Ref.\cite
{fyodorov} is exactly the same (although numeration of elements of the
matrix $Q$ is somewhat different) as the unitary version of $0D$ $\sigma $%
-model, Eq. (\ref{e1}). The pre-exponential is different but this is natural
because another (less direct) way of calculating the function $P\left(
\epsilon ,y\right) $ was used.

The same form of the $\sigma $-model obtained for these two different models
shows that the directed disordered model with broken time-reversal
invariance in a finite volume is equivalent to the model of weakly
non-Hermitian matrices. Apparently, the same equivalence holds between the
time reversal invariant model of disorder and models of weakly non-symmetric
real matrices. However, it is relevant to emphasize that not every
non-Hermitian Hamiltonian corresponds to the models of non-Hermitian or
non-symmetric real matrices. For example, models of open chaotic billiards
are described by Hamiltonians with additional imaginary terms (see, e.g.\cite
{ver,efetov}). These Hamiltonians do not seem to be equivalent to the random
matrix models of Ref.\cite{fyodorov}.

Now let us show how explicit calculations in Eqs. (\ref{e1},\ref{e3}) can be
performed. First of all one should choose a proper parametrization of the
supermatrices $Q$. The authors of Ref.\cite{fyodorov} used the
parametrization of Ref.\cite{efetov0} (``standard parametrization'' in
terminology of Ref.\cite{efetov}). This parametrization has been used for
solving many interesting problems. However, due to presence of the new terms
in the free energy $F\left[ Q\right] $, Eq. (\ref{e1}), this parametrization
is not as convenient as before\cite{efetov} because now $F\left[ Q\right] $
would contain not only the ``eigenvalues'' $\hat \theta $ but also many
other variables.

As concerns the unitary ensemble, the computation of the function $P\left(
\epsilon ,y\right) $is still possible although is very lengthy\cite{fyodorov}%
. At the same time, calculations for the orthogonal case using the standard
parametrization do not seem to be possible at all due to unsurmountable
technical problems.

Fortunately, one more parametrization is possible that is perfectly suitable
for the present problem. To some extent it resembles the parametrization
used to study the crossover between the orthogonal and unitary ensembles\cite
{altland,efetov}. Of course, it should be written for the orthogonal and
unitary ensembles in a different way but the main structure is the same. Let
us show in this Section how the function $P\left( \epsilon ,y\right) $ can
obtained for the unitary ensemble using this new parametrization (It can be
named ``non-Hermitian parametrization''). The orthogonal ensemble will be
considered in the next Section.

The supermatrix $Q$ in the non-Hermitian parametrization is written in the
form 
\begin{equation}
\label{e7}Q=TQ_0\bar T 
\end{equation}
where $T$ should be chosen to satisfy the relations $\left[ T,\Lambda
_1\right] =0$, $\bar TT=1$. The bar stands for the ``charge conjugation''
defined in Refs.\cite{efetov0,efetov}. It is clear that with such a choice
the function $F\left[ Q\right] $ would depend on $Q_0$ only (for the unitary
ensemble one has also $\left[ Q_0,\tau _3\right] =0$).

The central part $Q_0$ in Eq. (\ref{e7}) is taken in the form 
\begin{equation}
\label{e8}Q_0=\left( 
\begin{array}{cc}
\cos \hat \varphi & -\tau _3\sin 
\hat \varphi \\ -\tau _3\sin \hat \varphi & -\cos \hat \varphi 
\end{array}
\right) ,\qquad \hat \varphi =\left( 
\begin{array}{cc}
\varphi & 0 \\ 
0 & i\chi 
\end{array}
\right) 
\end{equation}
while the supermatrix $T$ can be chosen as 
\begin{equation}
\label{e9}T=\left( 
\begin{array}{cc}
u & 0 \\ 
0 & u 
\end{array}
\right) \left( 
\begin{array}{cc}
\cos \left( \hat \theta /2\right) & -i\sin \left( 
\hat \theta /2\right) \\ -i\sin \left( \hat \theta /2\right) & \cos \left( 
\hat \theta /2\right) 
\end{array}
\right) \left( 
\begin{array}{cc}
v & 0 \\ 
0 & v 
\end{array}
\right) 
\end{equation}
The supermatrices $\hat \theta $, $u$, $v$ are equal to 
\begin{equation}
\label{e10}\hat \theta =\left( 
\begin{array}{cc}
\theta & 0 \\ 
0 & i\theta _1 
\end{array}
\right) 
\end{equation}
$$
u=\left( 
\begin{array}{cc}
1-2\eta \bar \eta & 2\eta \\ 
-2\bar \eta & 1-2\bar \eta \eta 
\end{array}
\right) ,\qquad v=\left( 
\begin{array}{cc}
1-2\kappa \bar \kappa & 2\kappa \\ 
-2\bar \kappa & 1-2\bar \kappa \kappa 
\end{array}
\right) 
$$
The $2\times 2$ matrices $\varphi $, $\chi $, $\theta $, and $\theta _1$ are
proportional to the unit matrix, the matrices $\eta $, $\kappa $ are 
\begin{equation}
\label{e11}\eta =\left( 
\begin{array}{cc}
\eta & 0 \\ 
0 & -\eta ^{*} 
\end{array}
\right) ,\qquad \kappa =\left( 
\begin{array}{cc}
\kappa & 0 \\ 
0 & -\kappa ^{*} 
\end{array}
\right) 
\end{equation}
where $\eta $, $\eta ^{*}$, $\kappa $, and $\kappa ^{*}$ are anticommuting
variables. The conjugate matrices $\bar \eta $ and $\bar \kappa $ are the
same as in Refs.\cite{efetov0,efetov}. To understand better the structure of
the supermatrix $Q$ given by Eqs. (\ref{e7}-\ref{e11}) it is instructive to
write it neglecting all Grassmann variables. Then, one can write separately
the compact and noncompact sectors. The compact sector takes the form 
\begin{equation}
\label{e12}\left( 
\begin{array}{cc}
\cos \theta \cos \varphi & -\tau _3\sin \varphi +i\sin \theta \cos \varphi
\\ 
-\tau _3\sin \varphi -i\sin \theta \cos \varphi & -\cos \theta \cos \varphi 
\end{array}
\right) 
\end{equation}
whereas the noncompact sector is written as 
\begin{equation}
\label{e13}\left( 
\begin{array}{cc}
\cosh \theta _1\cos \chi & -i\tau _3\sinh \chi -\sinh \theta _1\cosh \chi \\ 
-i\tau _3\sinh \chi +\sinh \theta _1\cosh \chi & -\cosh \theta _1\cos \chi 
\end{array}
\right) 
\end{equation}
Comparing Eqs. (\ref{e12},\ref{e13}) with the corresponding expressions for
the supermatrix $Q$ in the standard parametrization\cite{efetov0,efetov} one
can understand that in order to specify the supermatrix $Q$ unambiguously
the following inequalities should be imposed 
\begin{equation}
\label{e14}-\infty <\chi <\infty ,\;-\infty <\theta _1<\infty ,\;-\pi
<\theta <\pi ,\;-\pi /2<\varphi <\pi /2 
\end{equation}
To start computation with the parametrization, Eqs. (\ref{e7}-\ref{e11}),
one should derive first the proper Jacobian. The derivation is presented in
the Appendix. The final result for the elementary volume $\left[ dQ\right] $
reads 
\begin{equation}
\label{e15}\left[ dQ\right] =J_\varphi J_\theta dR_BdR_F,\;dR_B=d\theta
d\theta _1d\varphi d\chi ,\;dR_F=d\eta d\eta ^{*}d\kappa d\kappa 
\end{equation}
where 
\begin{equation}
\label{e16}J_\varphi =\frac 1{8\pi }\frac{\cos \varphi \cosh \chi }{\left(
\sinh \chi +i\sin \varphi \right) ^2} 
\end{equation}
\begin{equation}
\label{e16a}J_\theta =\frac 1{32\pi }\frac 1{\sinh ^2\frac 12\left( \theta
_1+i\theta \right) } 
\end{equation}
Substututing Eqs. (\ref{e7}-\ref{e11}) for $Q$ in Eq. (\ref{e1}) one can
rewrite the function $F\left[ Q\right] $ in the limit $\tilde \gamma
\rightarrow 0$ as 
\begin{equation}
\label{e17}F\left[ Q\right] =a^2\left( \sinh ^2\chi +\sin ^2\varphi \right)
-ix\left( \sinh \chi +i\sin \varphi \right) 
\end{equation}
(The limit $\tilde \gamma \rightarrow 0$ is taken in the beginning of the
calculations because in the present parametrization this does not lead to
additional convergence problems). The function $F\left[ Q\right] $, Eq. (\ref
{e17}), does not contain the anticommuting variables and therefore one can
easily integrate over the supermatrix $u$. Writing in Eq. (\ref{e3}) the
supermatrix $Q$, Eqs. (\ref{e7}-\ref{e9}), as 
\begin{equation}
\label{e18}Q=u\tilde Q\bar u 
\end{equation}
with $u$ from Eq. (\ref{e10}) and integrating over $\eta $, $\eta ^{*}$ one
obtains%
$$
P\left( \epsilon ,y\right) =\frac{\pi \nu }{4\Delta }\int \left( STr\left(
\tau _3\Lambda _1\tilde Q\right) \right) ^2\exp \left( -F\left[ \tilde Q%
\right] \right) d\tilde Q 
$$
\begin{equation}
\label{e19}=\frac{4\pi \nu }\Delta \frac{d^2}{dx^2}\int \exp \left( -F\left[
Q\right] \right) d\tilde Q 
\end{equation}
where the elementary volume $\left[ d\tilde Q\right] $ differs from $\left[
dQ\right] $ by the replacement $dR_F\rightarrow d\tilde R_F=d\kappa d\kappa
^{*}$ and $F\left[ Q\right] $ is given by Eq. (\ref{e17}). Although Eq. (\ref
{e19}) is quite simple, one more difficulty should be overcome. The problem
is that the integrand in Eq. (\ref{e19}) does not contain the variables $%
\kappa $, $\kappa ^{*}$ and, at first glance, the integral must turn to
zero. However, the Jacobian $J_\theta $, Eq. (\ref{e16a}), is singular for $%
\theta ,\theta _1\rightarrow 0$ and this singularity is not compensated by
the integrand. So, one obtains an expression of the type $0\times \infty ,$
which is a usual phenomenon. Different procedures how to make the integral
well defined have been worked out (for a detailed discussion see\cite{efetov}%
). The simplest way is to rewrite Eq. (\ref{e3}) as 
\begin{equation}
\label{e20}P\left( \epsilon ,y\right) =P_m\left( \epsilon ,y\right) -\frac{%
\pi \nu }{4\Delta }\int A\left[ Q\right] \left( \exp \left( -F\left[
Q\right] \right) -\exp \left( -F_m\left[ Q\right] \right) \right) dQ 
\end{equation}
where 
\begin{equation}
\label{e21}P_m\left( \epsilon ,y\right) =-\frac{\pi \nu }{4\Delta }\int
A\left[ Q\right] \exp \left( -F_m\left[ Q\right] \right) dQ, 
\end{equation}

$$
F_m\left[ Q\right] -F\left[ Q\right] =-mSTr\left( T\Lambda \bar T\Lambda
\right) \equiv -mSTr\left( \Omega \Lambda \right) 
$$
The supermatrix $\Omega $ in Eq. (\ref{e21}) can be chosen as 
\begin{equation}
\label{e21b}\Omega =\tilde T\Lambda \overline{\tilde T},\qquad T=u\tilde T 
\end{equation}
The parameter $c$ in Eqs. (\ref{e20}, \ref{e21}) is arbitrary. Using Eq. (%
\ref{e9}) we see that 
\begin{equation}
\label{e21a}-STr\left( T\Lambda \bar T\Lambda \right) =4\left( \cosh \theta
_1-\cos \theta \right) 
\end{equation}
and thus, the singularity at $\theta _1=\theta =0$ coming from the Jacobian
in Eq. (\ref{e20}) is compensated by the integrand. After integration over $%
\eta ,\eta ^{*}$ the integrand does not contain the anticommuting variables $%
\kappa ,\kappa ^{*}$ and the integral vanishes. Therefore, the function $%
P_m\left( \epsilon ,y\right) $, Eq. (\ref{e21}), does not depend on $m$ and
one can calculate the integral in the limit $m\rightarrow \infty $. In this
limit only small deviations of the supermatrix $\Omega $ from $\Lambda $ are
essential. Using the representation, 
\begin{equation}
\label{e22}\Omega =\Lambda \left( 1+iW\right) \left( 1-iW\right) ^{-1},\text{%
\ }W=\left( 
\begin{array}{cc}
0 & B \\ 
B & 0 
\end{array}
\right) ,\;B=\left( 
\begin{array}{cc}
a & \sigma \\ 
\bar \sigma & ib 
\end{array}
\right) 
\end{equation}
expanding $\Omega $ in $W$ up to quadratic terms and calculating the
Jacobian in this approximation one can see that in the limit $m\rightarrow
\infty $ 
\begin{equation}
\label{e23}\int \exp \left( -mSTr\left( \Omega \Lambda \right) \right)
d\Omega =1 
\end{equation}
The supermatrix $\tilde T$ can also be represented through $W$ and
calculating the corresponding Jacobian one may expand up to quadratic in $W$
terms. As concerns $Q$ in the other terms in the integrand in Eq. (\ref{e21}%
), one should replace in the limit $m\rightarrow \infty $ the supermatrices $%
\tilde T$ by $1$. One can check also that now the Jacobian of the
transformation from the matrices $\tilde T$ and $u$ to $T$ equals to $-1$
and not to $J_\theta $ as it was with the initial parametrization for $T$,
Eq. (\ref{e9}).

So, calculating the integral, Eq. (\ref{e3}), one should replace the
supermatrix $\tilde T$ in the integrand by $1$. In the elementary volume $%
\left[ dQ\right] $, Eq. (\ref{e15}), one should omit the multiplier $%
J_\theta d\kappa d\kappa ^{*}$ and change the sign of the rest.

As a result of all these manipulations one comes to the following expression
for the function $P\left( \epsilon ,y\right) $ 
\begin{equation}
\label{e24}P\left( \epsilon ,y\right) =-\frac{\pi \nu }{4\Delta }\int \left(
STr\left( \tau _3\Lambda _1Q_0\right) \right) ^2\exp \left( -F\left[
Q_0\right] \right) J_\varphi d\varphi d\chi 
\end{equation}
with $Q_0$ from Eq. (\ref{e8}) and $J_\varphi $ from Eq. (\ref{e16}). The
function $F\left[ Q_0\right] $ is given by the R.H.S. of Eq. (\ref{e17}).
The limits of integration over $\varphi $ and $\chi $ are determined in Eqs.
(\ref{e14}).

The further calculation in Eq. (\ref{e24}) is very simple because the
function in the pre-exponential is proportional to $J_\varphi ^{-1}$.
Changing the variables of integration $z=\sinh \chi $, $t=\sin \varphi $,
one is to calculate a Gaussian integral over $z$, and the final expression
takes the form 
\begin{equation}
\label{e25}P\left( \epsilon ,y\right) =\frac{\nu \sqrt{\pi }}{a\Delta }\exp
\left( -\frac{x^2}{4a^2}\right) \int_0^1\cosh xt\exp \left( -a^2t^2\right)
dt 
\end{equation}
The function $P\left( \epsilon ,y\right) $ is properly normalized and one
obtains using Eq. (\ref{e2}) 
\begin{equation}
\label{e25a}\int P\left( \epsilon ,y\right) dy=1 
\end{equation}
The density of complex eigenvalues $P\left( \epsilon ,y\right) $, Eq. (\ref
{e25}), agrees precisely with the corresponding function for weakly
non-Hermitian random matrices obtained in Ref.\cite{fyodorov}. The
parameters $a$ and $\Delta $ are related in this case to the parameters in
Eq. (\ref{e5}, \ref{e6}) as 
\begin{equation}
\label{e26}a=\sqrt{2}\pi J\nu \left( \epsilon \right) \alpha ,\;\Delta
=\left( \nu \left( \epsilon \right) N\right) ^{-1},\text{\ }\nu \left(
\epsilon \right) =\left( 2\pi J\right) ^{-1}\sqrt{4-\left( \epsilon
/J\right) ^2} 
\end{equation}
and $x=2\pi \nu \left( \epsilon \right) yN$.

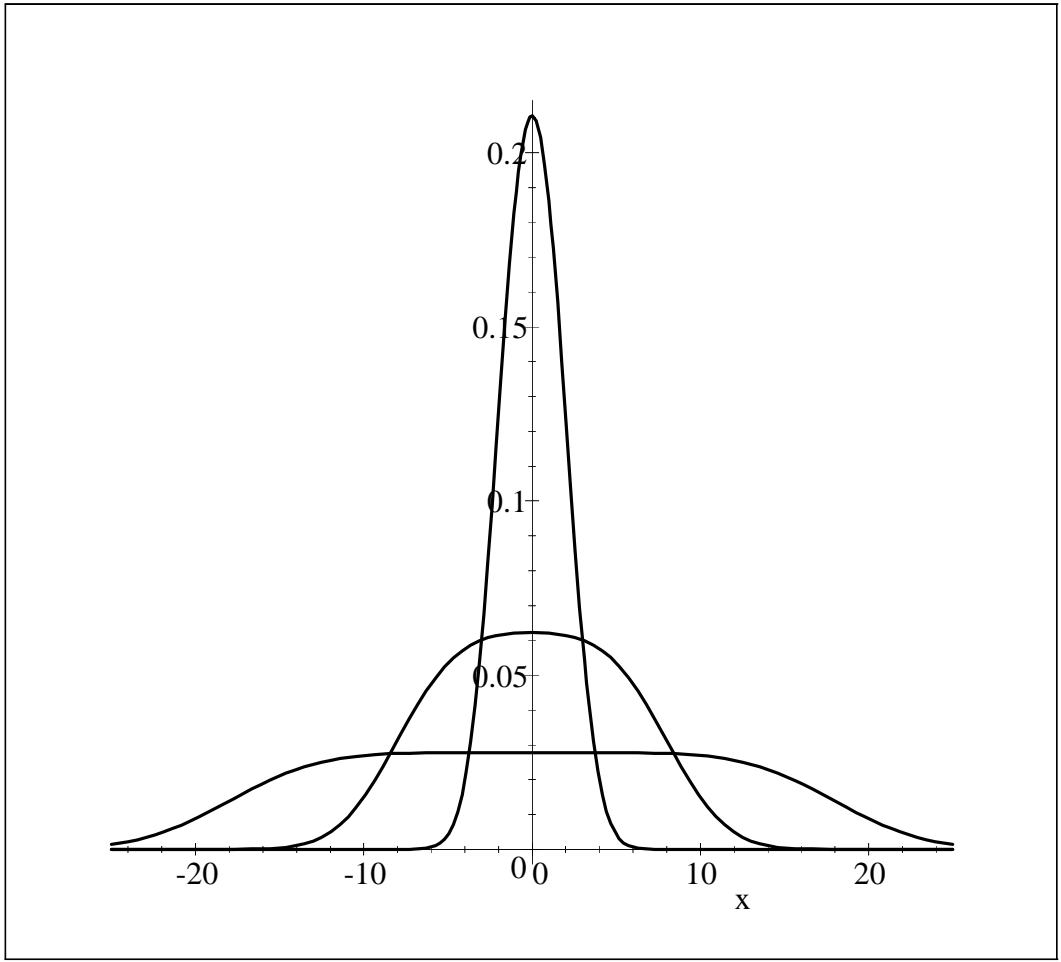
\begin{figure}
\centering
\vspace{2cm}
\caption{The density of complex eigenenergies $P\left( \epsilon ,y\right) $ 
for the unitary ensemble as a function of the imaginary part ($x=2\pi y/\Delta $)  for $a=1,2,3$}
\label{fig1}
\end{figure}

The agreement can serve as a proof of the equivalence between the directed
disorder models in a finite volume (with broken time reversal invariance)
and the models of non-Hermitian matrices defined by Eqs. (\ref{e5}, \ref{e6}%
). The function $P\left( \epsilon ,y\right) $ is represented in Fig.\ref
{fig1}. Its basic properties have been discussed in Ref.\cite{fyodorov}.

The density of complex eigenvalues is a smooth function at any finite $a$,
which means that any finite non-Hermiticity smears all eigenenergies making
them complex. The probability of real eigenvalues is negligible. For $a\gg 1$
the integral in Eq. (\ref{e25}) can be calculated analytically using the
saddle-point method. In the interval $\left| x\right| <2a^2$ the integrand
as a function of $t$ has a sharp maximum in the domain of the integration
and the integral can be extended to infinity. For $\left| x\right| >2a^2$
the function $P$ decays fast. As a result one obtains 
\begin{equation}
\label{e27}P\left( \epsilon ,y\right) \simeq \frac{\pi \nu \left( \epsilon
\right) }{2a^2\Delta }\left\{ 
\begin{array}{c}
1,\;\left| x\right| <2a^2 \\ 
0,\;\left| x\right| >2a^2 
\end{array}
\right. 
\end{equation}
Eq. (\ref{e27}) shows that for $a\gg 1$ the density of imaginary parts $y$
of eigenvalues at a fixed real part is homogeneous in the interval $x\in
\left( -2a^2,2a^2\right) $. Using Eq. (\ref{e26}) for $\nu \left( \epsilon
\right) $ and $a$ we can rewrite the result expressed by Eq. (\ref{e27}) in
terms of distribution of eigenvalues in the complex plane. In such a
formulation, Eq. (\ref{e27}) means that the complex eigenvalues are
distributed homogeneously within the ellipse 
\begin{equation}
\label{e28}\left( \frac \epsilon {2J}\right) ^2+\left( \frac y{2Jv}\right)
^2=1,\qquad v=\alpha N^{-1/2} 
\end{equation}
This is the ``elliptic law'' found in Refs.\cite{girko,sommers}, which is
natural because the limit $a\gg 1$ should correspond to a ``strong''
non-Hermiticity. At the same time, it is clear the elliptic law is model
dependent. For the models of disorder considered in the present paper the
density of complex states essentially depends on $y$ only.

In the opposite limit $a\ll 1$ the density of complex states $P\left(
\epsilon ,y\right) $ takes the form 
\begin{equation}
\label{e29}P\left( \epsilon ,y\right) \simeq \frac{\nu \sqrt{\pi }}{a\Delta }%
\exp \left( -\frac{x^2}{4a^2}\right) 
\end{equation}
The Gaussian form of the function $P$ can be easily understood starting from
the random matrix model, Eqs. (\ref{e5}, \ref{e6}). The function $P\left(
\epsilon ,y\right) $ can be written as 
\begin{equation}
\label{e30}P\left( \epsilon ,y\right) =N^{-1}\sum_{n=1}^N\left\langle \delta
\left( \epsilon -\epsilon _n^{\prime }\right) \delta \left( y-\epsilon
_n^{\prime \prime }\right) \right\rangle 
\end{equation}
$$
=\frac 1{2\pi N}\sum_{n=1}^N\int_{-\infty }^\infty dke^{iky}\left\langle
\delta \left( \epsilon -\epsilon _n^{\prime }\right) \exp \left( -ik\epsilon
_n^{\prime \prime }\right) \right\rangle 
$$
Where the angle brackets $\left\langle ...\right\rangle $ stand for the
averaging over the matrices $\hat A$ and $\hat B$, Eq. (\ref{e6}). In the
limit of small $\alpha $ the imaginary part $\epsilon _n^{\prime \prime }$
can be obtained using the standard perturbation theory. In the first order
one has 
\begin{equation}
\label{e31}\epsilon _m^{\prime \prime }=\vec \phi _m^{*}\hat B\vec \phi _m 
\end{equation}
where $\vec \phi _m$ is the eigenvector of the matrix $\hat A$ corresponding
to the eigenvalue $\epsilon _m^{\prime }$. Substituting Eq. (\ref{e31}) into
Eq. (\ref{e30}) one can immediately average over the matrix $\hat B$. Using
the orthogonality of the eigenvectors $\vec \phi _m$ one can write the
result of the averaging as 
\begin{equation}
\label{e32}P\left( \epsilon ,y\right) =\frac 1{2\pi N}\sum_{n=1}^N\int_{-%
\infty }^\infty dke^{iky}\left\langle \delta \left( \epsilon -\epsilon
_n^{\prime }\right) \exp \left[ -\frac 12\left( \frac{\alpha kJ}N\right)
^2\right] \right\rangle _A 
\end{equation}
where $\left\langle ...\right\rangle $ stands for averaging over $\hat A$.
Integrating over and using Eq. (\ref{e26}) one comes to Eq. (\ref{e29}). As
concerns the models of disorder, Eqs. (\ref{a1}-\ref{a2}), even the
asymptotics, Eqs. (\ref{e27}, \ref{e32}), have not been known before and it
not clear how to reproduce them using simple arguments.

Are the result obtained in this Section general and one cannot expect
anything new for the orthogonal ensemble? Of course, there is no reason to
hope that Eq. (\ref{e25}) describes the orthogonal ensemble as well but are
the asymptotics in the limits $a\gg 1$ and $a\ll 1$, Eqs. (\ref{e27}-\ref
{e29}) still correct?

The orthogonal ensemble of random matrices can be introduced again by Eqs. (%
\ref{e5}, \ref{e6}) but now the matrices $\hat A$ and $\hat B$ should be
real symmetric and antisymmetric, respectively. One should also make in Eq. (%
\ref{e5}) the replacement $\alpha \rightarrow -i\alpha $. As concerns the
asymptotics in the limit $a\gg 1$ the same elliptic law as in Eq. (\ref{e28}%
) has been recovered\cite{leh}. At the same time, one can expect completely
different behavior for $a\ll 1$. This can be seen easily from the fact that
the first order of the perturbation theory corresponding to Eq. (\ref{e31})
gives zero and one cannot derive Eq. (\ref{e29}) as before. In fact, the
density of complex eigenvalues $P\left( \epsilon ,y\right) $ is singular at $%
y=0$. Study of the orthogonal ensemble is presented in the next Section.

\section{ Density of complex eigenvalues in a limited volume: orthogonal
ensemble}

To compute the density of complex eigenvalues $P\left( \epsilon ,y\right) $
for the orthogonal ensemble one can start as previously from Eqs. (\ref{e1}-%
\ref{e4}) but now one should use supermatrices $Q$ with the structure
corresponding to this case. As has been mentioned, the presence in Eq. (\ref
{e1}) of the new term with the matrix $\Lambda _1$ makes the calculation
very difficult even for the unitary ensemble and hardly feasible at all for
the orthogonal one. So, as in the preceding Section a new parametrization
for $Q$ should be designed.

Let us write the supermatrix $Q$ in the form 
\begin{equation}
\label{o1}Q=ZQ_0\bar Z,\qquad Z=TY 
\end{equation}
with the supermatrices $Q_0$ and $T$ specified by Eqs. (\ref{e8}-\ref{e11})
and choose the supermatrix $Y$ as follows 
\begin{equation}
\label{o2}Y=Y_0RS,\;Y_0=Y_3Y_2Y_1 
\end{equation}
The supermatrix $Y_1$ entering Eq. (\ref{o2}) is 
\begin{equation}
\label{o3}Y_1=\left( 
\begin{array}{cc}
\hat w & 0 \\ 
0 & \hat w 
\end{array}
\right) ,\;\hat w=\left( 
\begin{array}{cc}
w & 0 \\ 
0 & 1 
\end{array}
\right) ,\;w=\left( 
\begin{array}{cc}
\cos \left( \mu /2\right) & -\sin \left( \mu /2\right) \\ 
\sin \left( \mu /2\right) & \cos \left( \mu /2\right) 
\end{array}
\right) 
\end{equation}
The supermatrix $Y_2$ is equal to 
\begin{equation}
\label{o4}Y_2=\left( 
\begin{array}{cc}
\cos \left( \hat \theta _2/2\right) & -i\sin \left( 
\hat \theta _2/2\right) \\ -i\sin \left( \hat \theta _2/2\right) & \cos
\left( \hat \theta _2/2\right) 
\end{array}
\right) ,\;\hat \theta _2=\left( 
\begin{array}{cc}
0 & 0 \\ 
0 & i\theta _2\tau _1 
\end{array}
\right) ,\;\tau _1=\left( 
\begin{array}{cc}
0 & 1 \\ 
1 & 0 
\end{array}
\right) 
\end{equation}
The supermatrix $Y_3$ is 
\begin{equation}
\label{o5}Y_3=\left( 
\begin{array}{cc}
\exp \left( i\hat \beta /2\right) & 0 \\ 
0 & \exp \left( i\hat \beta /2\right) 
\end{array}
\right) ,\qquad \hat \beta =\left( 
\begin{array}{cc}
\beta \tau _3 & 0 \\ 
0 & \beta _1\tau _3 
\end{array}
\right) ,\;\tau _3=\left( 
\begin{array}{cc}
1 & 0 \\ 
0 & -1 
\end{array}
\right) 
\end{equation}
The supermatrices $R$ and $S$ contain remaining Grassmann variables and are
written as 
\begin{equation}
\label{o6}R=\left( 
\begin{array}{cc}
\hat R & 0 \\ 
0 & \hat R 
\end{array}
\right) ,\;\hat R=\left( 
\begin{array}{cc}
1-2\rho \bar \rho & 2\rho \\ 
-2\bar \rho & 1+2\rho \bar \rho 
\end{array}
\right) ,\;\rho =\left( 
\begin{array}{cc}
\rho & 0 \\ 
0 & -\rho ^{*} 
\end{array}
\right) 
\end{equation}
and 
\begin{equation}
\label{o7}S=\left( 
\begin{array}{cc}
1-2\hat \sigma ^2 & 2i 
\hat \sigma \\ 2i\hat \sigma & 1-2\hat \sigma ^2 
\end{array}
\right) ,\qquad \hat \sigma =\left( 
\begin{array}{cc}
0 & \sigma \\ 
\bar \sigma & 0 
\end{array}
\right) ,\;\sigma =\left( 
\begin{array}{cc}
\sigma & 0 \\ 
0 & -\sigma ^{*} 
\end{array}
\right) 
\end{equation}
where $\bar \rho $ and $\bar \sigma $ are conjugate to $\rho $ and $\sigma $.

The parametrization for $Y$, Eqs. (\ref{o2}-\ref{o7}), is chosen in such a
way that $\left[ Y,\Lambda _1\right] =0$. To specify the supermatrix $Q$
unambiguously one should restrict variations of the variables by certain
intervals. This can be done as the preceding Section by comparing the
bosonic ``skeleton'' of $Q$ written in the parametrization, Eq. (\ref{o1}-%
\ref{o7}), (let us called it ``non-symmetric parametrization'') with the
standard parametrization of Refs.\cite{efetov0,efetov}. As a result one can
write the following inequalities 
\begin{equation}
\label{o8}
\begin{array}{c}
0<\chi <\infty ,\;-\pi /2<\varphi <\pi /2,\;-\infty <\theta _1<\infty
,\;-\pi <\theta <\pi \\ 
0<\theta _2<\infty ,\;0<\mu <\pi ,\;0<\beta <\pi ,\;0<\beta _1<2\pi 
\end{array}
\end{equation}
The next step is to calculate the Jacobian. The derivation is presented in
the Appendix and the final result for the elementary volume $\left[
dQ\right] $ is 
\begin{equation}
\label{o9}\left[ dQ\right] =J_\varphi J_\theta J_\mu
J_cdR_BdR_FdR_{1B}dR_{1F} 
\end{equation}
In Eq. (\ref{o9}), $J_\varphi ,$ $J_\theta ,dR_B$ and $dR_F$ are given by
Eqs. (\ref{e15}-\ref{e16a}). The additional quantities entering Eq. (\ref{o9}%
) are equal to 
\begin{equation}
\label{o10}J_\mu =\frac 1{2^8\pi ^2}\frac{\sinh \theta _2\sin \mu }{\left(
\cosh \theta _2-\cos \mu \right) ^2} 
\end{equation}
\begin{equation}
\label{o11}J_c=\frac{4\sin ^2\varphi }{\left( \sinh \chi -i\sin \varphi
\right) ^2} 
\end{equation}
and 
\begin{equation}
\label{o12}dR_{1B}=d\mu d\theta _2d\beta d\beta _1,\;dR_{1F}=d\sigma d\sigma
^{*}d\rho d\rho ^{*} 
\end{equation}
The free energy $F\left[ Q\right] $, Eq. (\ref{e1}), takes in the limit $%
\gamma \rightarrow 0$ the following form

\begin{equation}
\label{o13}F\left[ Q\right] =a^2\left( \sin ^2\varphi +\sinh ^2\chi \right)
+x[\left( \cos \mu \sin \varphi -i\cosh \theta _2\sinh \chi \right) 
\end{equation}

$$
+4\left( \sigma \sigma ^{*}+\rho \rho ^{*}\right) \left( \cosh \theta
_2-\cos \mu \right) \left( \sin \varphi -i\sinh \chi \right) ] 
$$
The non-symmetric parametrization given by Eqs. (\ref{o1}-\ref{o12}) looks
rather complicated. The calculation of the Jacobian is most lengthy but this
has to be done only once. At the same time, the Jacobian does not contain
Grassmann variables and the free energy $F\left[ Q\right] $, Eq. (\ref{o13}%
), is simple enough. Moreover, the supermatrix $Q$ can be written as in the
preceding Section in the form of Eq. (\ref{e18}) (although the supermatrix $%
\tilde Q$ is now different from that for the unitary ensemble). This allows
to integrate first over the matrix $u$ and obtain Eq. (\ref{e19}).

Further simplifications come from the fact that as previously one obtains an
uncertainty of the type $0\times \infty $ because the integrand in Eq. (\ref
{e19}) does not contain the variables $\kappa ,\kappa ^{*}$ whereas the
Jacobians $J_\theta $, Eq. (\ref{e16a}), and $J_\mu $, Eq. (\ref{o10}), are
singular at $\theta ,\theta _1,\theta _2,\mu \rightarrow 0$. We have seen in
the preceding Section that the uncertainties can be rather easily avoided
and, as a result, one obtains a more simple integral. The ``regularization''
procedure, Eqs. (\ref{e20}-\ref{e23}), led to the integral, Eq. (\ref{e24}),
that contained the variables $\varphi $ and $\chi $ only.

Similar transformations can be performed for the orthogonal ensemble.
Proceeding as for the unitary ensemble let us introduce the function $%
F_{mn}\left[ Q\right] $%
\begin{equation}
\label{o14}F_{mn}=F\left[ Q\right] -mSTr\left( \Lambda T\Lambda \bar T%
\right) -nSTr\left( \tau _3Y\tau _3\bar Y\right) 
\end{equation}
The second term in Eq. (\ref{o14}) can also be written in the form of Eq. (%
\ref{e21a}). Using Eqs. (\ref{o2}-\ref{o7}) we can write the third term as 
\begin{equation}
\label{o15}-nSTr\left( \tau _3Y\tau _3\bar Y\right) =4n\left( \cosh \theta
_2-\cos \mu \right) 
\end{equation}
In analogy with the transformation of the integrand in Eqs. (\ref{e20}, \ref
{e21}), we can represent $\exp \left( -F\left[ Q\right] \right) $ as follows 
\begin{equation}
\label{o16}e^{-F}=e^{-F_{mn}}+e^{-F_{m0}}\left( 1-e^{-F^{\left( n\right)
}}\right) +e^{-F_{0n}}\left( 1-e^{-F^{\left( m\right) }}\right)
+e^{-F}\left( 1-e^{-F^{\left( m\right) }}\right) \left( 1-e^{-F^{\left(
n\right) }}\right) 
\end{equation}
where%
$$
F^{\left( m\right) }=F_{mn}-F_{0n},\;F^{\left( n\right) }=F_{mn}-F_{m0} 
$$
The parameters $m$ and $n$ in Eqs. (\ref{o15}, \ref{o16}) are arbitrary.
Therefore, substituting Eq. (\ref{o16}) into Eq. (\ref{e3}) we can take the
limit $m,n\rightarrow \infty $. The contribution coming from the last term
in Eq. (\ref{o16}) vanishes because all singularities are compensated for
any $m$ and $n$ but the integrand does not contain the anticommuting
variables $\kappa ,\kappa ^{*}$. The limit $m\rightarrow \infty $ allows to
expand the supermatrix $\tilde T$, Eq. (\ref{e21b}), near $1$ (and the
supermatrix $\Omega $ near $\Lambda $). As has been explained in the
preceding Section, in the limit $m\rightarrow \infty $ one can replace $%
\tilde T\rightarrow 1$ everywhere in the integrand omitting simultaneously $%
J_\theta d\kappa \kappa ^{*}$ in the elementary volume $\left[ dQ\right] $.
The same is correct now and one should remove $J_\theta d\kappa d\kappa ^{*}$
from $\left[ dQ\right] $, Eq. (\ref{o9}) (changing the sign).

The other singularity at $\theta _2,\mu \rightarrow 0$ in the first and
third terms in Eq. (\ref{o16}) can be avoided in a similar way. In the limit 
$n\rightarrow \infty $ the supermatrix $Y,$ Eqs. (\ref{o2}-\ref{o7}), is
also close to $1$. To make an expansion in small deviations $Y$ from $1$ one
can use the following parametrization 
\begin{equation}
\label{o17}Y=\left( 1-iX\right) \left( 1+iX\right) ^{-1},\;X=\left( 
\begin{array}{cc}
i\hat A & \hat L \\ \hat L & i\hat A 
\end{array}
\right) 
\end{equation}
The blocks $\hat A$ and $\hat L$ satisfy the constraints $\bar A=-A$, $\bar L%
=L$, $\left\{ A,\tau _3\right\} =0$, $\left\{ L,\tau _3\right\} =0$, where $%
\left\{ ...\right\} $ is anticommutator. These blocks can be written in an
explicit form as 
\begin{equation}
\label{o18}\hat A=\left( 
\begin{array}{cc}
f & \xi \\ 
-\bar \xi & 0 
\end{array}
\right) ,\;\hat L=\left( 
\begin{array}{cc}
0 & \zeta \\ 
\bar \zeta & il 
\end{array}
\right) 
\end{equation}
where the $2\times 2$ matrices $f$ and $l$contain conventional complex
numbers $f$ and $l$, whereas $\xi $ and $\zeta $ consist of anticommuting
variables $\xi $ and $\zeta $. The explicit form of these matrices is 
\begin{equation}
\label{o19}f=\left( 
\begin{array}{cc}
0 & -f \\ 
f^{*} & 0 
\end{array}
\right) ,\;l=\left( 
\begin{array}{cc}
0 & l \\ 
l^{*} & 0 
\end{array}
\right) ,\;\zeta =\left( 
\begin{array}{cc}
0 & \zeta \\ 
-\zeta ^{*} & 0 
\end{array}
\right) 
\end{equation}
In Eq. (\ref{o19}), $l$ is an arbitrary complex number, while for $f$ one
should integrate over the domain $Imf>0$.

Substituting Eqs. (\ref{o17}-\ref{o19}) into Eq. (\ref{o14}) one should
expand the term $STr\left( \tau _3Y\tau _3\bar Y\right) $ up to quadratic
terms in $X$ and replace $Y$ by $1$ everywhere else in the integrand.
Calculating the Jacobian we can see that the factor $J_\mu dR_{1B}dR_{1F}$
should be replaced by $1$. Of course, this concerns only the first and the
third terms in Eq. (\ref{o16}) because the second term does not lead to any
singularity in the integrand at $\theta _2=\mu =0$. In fact, the
contribution from the third term in Eq. (\ref{o16}) is zero because it is
not singular at $\theta =\theta _1=0$ and does not contain the variables $%
\kappa ,\kappa ^{*}$. At the same time we understand what to do with the
singularity at $\theta _2=\mu =0$.

The result of this discussion can be formulated finally as follows. We
should replace Eq. (\ref{e3}) by 
\begin{equation}
\label{o20}P\left( \epsilon ,y\right) =P^{\left( 1\right) }\left( \epsilon
,y\right) +P^{\left( 2\right) }\left( \epsilon ,y\right) , 
\end{equation}

\begin{equation}
\label{o21}P^{\left( 1\right) }\left( \epsilon ,y\right) =-\frac{\pi \nu }{%
4\Delta }\lim _{m,n\rightarrow \infty }\int A\left[ Q\right] \exp \left(
-F_{mn}\left[ Q\right] \right) dQ 
\end{equation}

\begin{equation}
\label{o22}P^{\left( 2\right) }\left( \epsilon ,y\right) =-\frac{\pi \nu }{%
4\Delta }\lim _{m,n\rightarrow \infty }\int A\left[ Q\right] \left( \exp
\left( -F_{m0}\right) -\exp \left( -F_{mn}\right) \right) dQ 
\end{equation}
The integrand in Eq. (\ref{o21}) has both singularities. Therefore, one has
to replace everywhere in the integrand $\tilde T$ and $Y$ by $1$
simultaneously replacing $J_\theta J_\mu d\kappa d\kappa ^{*}dR_{1B}dR_{1F}$
in the elementary volume $\left[ dQ\right] $, Eq. (\ref{o9}) by $-1$. As
concerns Eq. (\ref{o22}) the integrand has only the singularity at $\theta
=\theta _1=0$ and one should replace by $1$ the supermatrix $\tilde T$ only.
In the elementary volume $J_\theta d\kappa d\kappa ^{*}$ should be replaced
by $-1$.

The subsequent manipulations are rather straightforward. Integrating over
the supermatrix $u$ one obtains for $P^{\left( 1\right) }\left( \epsilon
,y\right) $ and $P^{\left( 2\right) }\left( \epsilon ,y\right) $analogs of
Eq. (\ref{e19}). Then, the function $P^{\left( 1\right) }\left( \epsilon
,y\right) $ is expressed in terms of the integral over the variables $t=\sin
\varphi $ and $z=\sinh \chi $ 
\begin{equation}
\label{o23}P^{\left( 1\right) }\left( \epsilon ,y\right) =\frac \nu {4\Delta 
}\frac{d^2}{dx^2}\int e^{-a^2\left( t^2+z^2\right) -x\left( t-iz\right) }%
\frac{4t^2dtdz}{\left( t^2+z^2\right) ^2} 
\end{equation}
In the integral in Eq. (\ref{o22}) one has to integrate first over the
variables $\rho ,\rho ^{*},\sigma $, and $\sigma ^{*}$ and then, the
function $P^{\left( 2\right) }\left( \epsilon ,y\right) $ reduces to 
\begin{equation}
\label{o24}P^{\left( 2\right) }\left( \epsilon ,y\right) =\frac \nu {4\Delta 
}\frac{d^2}{dx^2}\int e^{-a^2\left( t^2+z^2\right) -x\left( t\omega
-i\lambda z\right) }\frac{\left( t-iz\right) ^2x^2t^2}{\left( t^2+x^2\right)
^2}dtdzd\omega d\lambda 
\end{equation}
where $\omega =\cos \mu $, and $\lambda =\cosh \theta _2$. The integration
in Eq. (\ref{o23}, \ref{o24}) is performed over $t$ and $z$ in the intervals 
$-1<t<1,$ $-\infty <z<\infty $ and over $\omega $ and $\lambda $ in the
intervals $-1<\omega <1$, $1<\lambda <\infty $.

The integration over $\omega $ and $\lambda $ in Eq. (\ref{o24}) can be
carried out immediately. However, to provide the convergence of the integral
over $\lambda $ one should shift the contour of integration over $z$ into
the complex plane $z\rightarrow z+i\delta sgn\left( x\right) $, where $%
\delta $ is an infinitesimal positive number and%
$$
sgn\left( x\right) =\left\{ 
\begin{array}{c}
1,\;x>0 \\ 
-1,\;x<0 
\end{array}
\right. 
$$
Intergrating over $\omega $ and $\lambda $ and adding Eqs. (\ref{o23}, \ref
{o24}) we obtain for $P\left( \epsilon ,y\right) $, Eq. (\ref{o20}) 
\begin{equation}
\label{o25}P\left( \epsilon ,y\right) =\frac \nu {4\Delta }\frac{d^2I\left(
x\right) }{dx^2}, 
\end{equation}
\begin{equation}
\label{o25a}I\left( x\right) =\int_{-1}^1\int_{-\infty }^\infty
e^{-a^2\left( t^2+z_{-}^2\right) }\left[ e^{-x\left( t-iz_{-}\right) }\left(
t+iz_{-}\right) ^2-e^{x\left( t+iz_{-}\right) }\left( t-iz_{-}\right)
^2\right] \frac t{iz_{-}}\frac{dtdz}{\left( t^2+z_{-}^2\right) ^2} 
\end{equation}
where $z_{-}=z+i\delta sgn\left( x\right) $.

It is clear from the form of the function $I\left( x\right) $ that it is
convenient to differentiate first over $x$ and then calculate the integral.
However, one should be careful performing this, at first glance trivial,
manipulation. The problem is that $z_{-}$ contains $x$, which can result in
an additional contribution.

To avoid lengthy calculations let us consider first the case when $x$ is
finite nonzero number. Then, the derivatives $dz_{-}/dx$ and $d^2z_{-}/dx^2$
vanish and one has to differentiate the exponentials only. Shifting the
contour of integration $z\rightarrow z+\frac{ix}{2a^2}$, which can be done
without crossing singularities in the complex plane and changing the new
variable $z$ as $z\rightarrow z/a$ one obtains 
\begin{equation}
\label{o26}P_c\left( \epsilon ,y\right) =\frac \nu {a\Delta }\exp \left( -%
\frac{x^2}{4a^2}\right) \int_0^1xt\sinh xt\exp \left( -a^2t^2\right)
dt\int_0^\infty \frac{\exp \left( -z^2\right) dz}{z^2+\frac{x^2}{4a^2}} 
\end{equation}
(the variables $x$ and $y$ are related through Eq. (\ref{e2})).

Eq. (\ref{o26}) holds for any finite $x$ but is it the final result? It
would be the final result it the density function were continuous at $x=0$.
As concerns the unitary ensemble, we already know that the function $P\left(
\epsilon ,y\right) $is continuous (see Eq. (\ref{e25})) but does the
continuity follow from a physical principle? In fact it does not and the
function $P\left( \epsilon ,y\right) $ for the unitary ensemble contains a $%
\delta $-function at $x=0$.

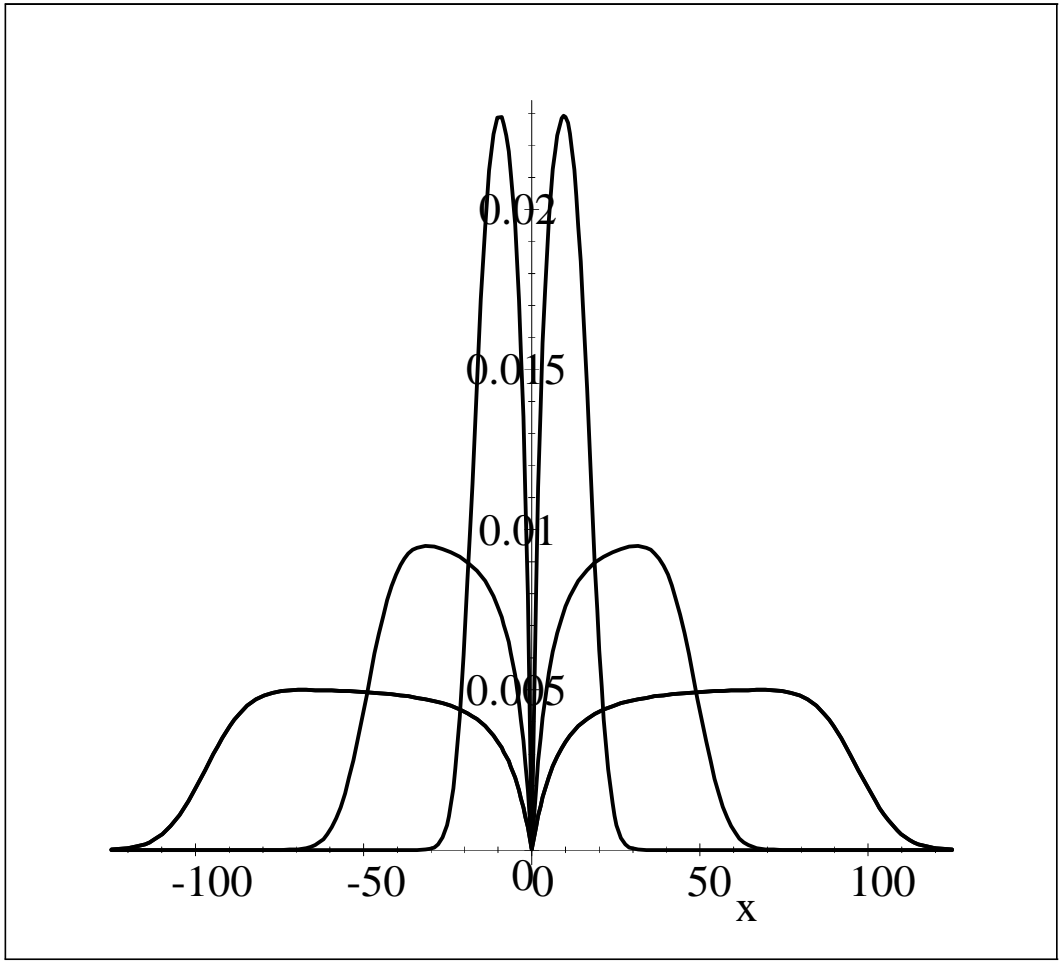
\begin{figure}
\centering
\vspace{2cm}
\caption{The density of complex eigenenergies $P_c\left( \epsilon ,y\right) $for the
orthogonal ensemble as a function of the imaginary part ($x=2\pi y/\Delta $)  for $a=3,5,7$}
\label{fig2}
\end{figure}

To extract the $\delta $-function let us expand the exponentials in the
integrand in Eq. (\ref{o25a}). In the first two orders one obtains 
\begin{equation}
\label{o27}P\left( \epsilon ,y\right) \simeq \frac \nu {2\Delta }\frac{d^2}{%
dx^2}\left[ \int_{-1}^1\int_{-\infty }^\infty t^2e^{-a^2\left(
t^2+z_{-}^2\right) }\left[ \frac 2{\left( t^2+z_{-}^2\right) ^2}-\frac x{%
iz_{-}}\frac 1{t^2+z_{-}^2}\right] dtdz\right] 
\end{equation}
The first term in the integrand in Eq. (\ref{o27}) has no singularities and
one can shift the contour of the integration over $z$ such the variables $%
z_{-}$ are replaced by $z$. Then, this part of the integrand does not
contain $x$ and the differentiation gives zero. The contribution involved
comes from the second term in the integrand. Writing $z_{-}^{-1}$ as%
$$
\frac 1{z_{-}}=\frac{z-i\delta sgn\left( x\right) }{z^2+\delta ^2} 
$$
one can represent the function $P\left( \epsilon ,y\right) $ for $%
x\rightarrow 0$ as 
$$
P\left( \epsilon ,y\right) _{x\rightarrow 0}=\frac \nu {2\Delta }\frac{d^2}{%
dx^2}\lim _{\delta \rightarrow 0}\int_{-1}^1\int_{-\infty }^\infty
e^{-a^2\left( t^2+z^2\right) }\frac{t^2\left| x\right| }{t^2+z^2}\frac \delta
{z^2+\delta ^2}dtdz 
$$
The integration over $z$ in the limit $\delta \rightarrow 0$ is elementary
and one obtains for the anomalous contribution $P_r\left( \epsilon ,y\right) 
$ the following expression 
\begin{equation}
\label{o28}P_r\left( \epsilon ,y\right) =\frac{2\pi \nu }\Delta \delta
\left( x\right) \int_0^1\exp \left( -a^2t^2\right) dt 
\end{equation}
Making some simple transformations in Eq. (\ref{o26}) the final result for
the density of complex eigenvalues $P\left( \epsilon ,y\right) $ can be
written as 
\begin{equation}
\label{o29}P\left( \epsilon ,y\right) =P_r\left( \epsilon ,y\right)
+P_c\left( \epsilon ,y\right) 
\end{equation}
where $P_r\left( \epsilon ,y\right) $ is given by Eq. (\ref{o28}) and $%
P_c\left( \epsilon ,y\right) $ equals 
\begin{equation}
\label{o30}P_c\left( \epsilon ,y\right) =\frac{2\pi \nu }\Delta \frac 12\Phi
\left( \frac{\left| x\right| }{2a}\right) \int_0^1t\sinh \left( \left|
x\right| t\right) \exp \left( -a^2t^2\right) dt 
\end{equation}
where $\Phi \left( v\right) =\frac 2{\sqrt{\pi }}\int_v^\infty \exp \left(
-u^2\right) du.$ It is not difficult to check that the function $P\left(
\epsilon ,y\right) $, Eqs. (\ref{o28}-\ref{o30}), satisfies the
normalization condition, Eq. (\ref{e25a}), and the singular part $P_r\left(
\epsilon ,y\right) $ gives an essential contribution that becomes small only
in the limit $a\rightarrow \infty $. The function $P_c\left( \epsilon
,y\right) $ is represented in Fig.\ref{fig2}.

The existence of the anomalous part $P_r\left( \epsilon ,y\right) $, Eq. (%
\ref{e28}), means that a finite fraction of all eigenvalues remains real for
any imaginary vector potential $h$ in the models of disorder, Eqs. (\ref{a1}%
, \ref{a2}) or degree of asymmetry $\alpha $ for the real random matrix
models. At the same time, the function $P_c\left( \epsilon ,y\right) $
decays when $y\rightarrow 0$, which corresponds to a vanishing probability
of eigenstates with small but nonzero imaginary parts.

In contrast to the unitary ensemble, the function $P\left( \epsilon
,y\right) $ for $a\ll 1$ can hardly be obtained from a perturbation theory.
Most of the eigenvalues are in this case real. In the opposite limit $a\gg 1$
one should distinguish between several regions. In the limit $\left|
x\right| \ll a$ the asymptotics is determined by the expression 
\begin{equation}
\label{o31}P_c\left( \epsilon ,y\right) \simeq \frac{\pi \nu }{2a^2\Delta }%
\frac{\sqrt{\pi }\left| x\right| }{2a} 
\end{equation}
showing a linear decay of the density as $\left| x\right| \rightarrow \infty 
$.

In the region $\left| x\right| \gg 2a$ the density of complex eigenvalues is
constant for $\left| x\right| <2a^2$ and falls off outside this interval.
Its value in this region is the same as in the unitary case, Eq. (\ref{e27}%
). This corresponds to the elliptic law, Eq. (\ref{e28}). For an ensemble of
strongly asymmetric real random matrices with a Gaussian distribution this
law has been proven in Ref.\cite{sommers,leh}. The authors of this
publication have also found numerically that the portion of real eigenvalues
for their ensemble decays as $N^{-1/2},$ where $N$ is the size of the
matrices. Apparently, this behavior corresponds to the $\delta $-functional
part $P_r\left( \epsilon ,y\right) $, Eq. (\ref{o28}), in the eigenvalue
density for the case of weak asymmetry (orthogonal analog of Eqs. (\ref{e6}, 
\ref{e6})).

\section{Discussion}

The results presented in the previous Sections demonstrate that the disorder
models with a direction are interesting and can be efficiently studied using
the supersymmetry technique. The $\sigma $-model derived, Eq. (\ref{a42}),
can be used in any dimension. It is relevant to emphasize that, as usual\cite
{efetov0,efetov}, the dimensionality is determined by the geometry of the
sample. So, the one-dimensional version of the $\sigma $-model corresponds
to a thick wire with a directed hopping. In the language of vortices in a
superconductor \cite{hatano} the $1D$ model can describe the vortices in a
slab with line defects and the magnetic field parallel to the surface. Such
a model is somewhat more realistic than a purely $1D$ model of Ref.\cite
{hatano}. The $2D$ $\sigma $-model is supposed to describe the vortices in a
bulk superconductor with line defects. In addition, one can imagine a
situation when the sample is long but has a small cross-section. If the line
defects are aligned in the longitudinal direction one comes to the $0D$ $%
\sigma $-model considered in the present paper.

Of course, the directed non-Hermitian Hamiltonians can arise not only from
the vortex model but also correspond to non-equilibrium processes. A very
interesting possibility is the directed hopping model, Eq. (\ref{a2}) that
can be considered as a quantum counterpart of the directed percolation model%
\cite{obukhov}. Applications to other physical systems that can be reduced
to models of a disorder with a direction also deserve an attention. The
problem of turbulence is one of most famous. The main features of the
turbulence are believed to be described by the Burgers equation\cite
{burgers,polyakov,chekhlov}. Reduction of the Burgers equation to a linear
equation allows to use well developed methods of disorder physics. A
similarity of the linear equation to equations used in study of problems of
directed polymers have already inspired application of the replica method to
study the problem of turbulence\cite{bouchaud}. Use of the supersymmetry for
the problems of the turbulence might be one more interesting direction of
research.

Leaving these interesting problems for future study let us summarize the
results obtained in the present work. The $\sigma $-model, Eq. (\ref{a42}),
differs from the $\sigma $-models used in the localization and mesoscopic
problems\cite{efetov0,efetov} by the term with the matrix $\Lambda _1$.
Although the Hamiltonians with the direction, Eqs. (\ref{a1}, \ref{a2}) can
be obtained from conventional Hermitian Hamiltonians in a magnetic field by
the formal replacement ${\bf A\rightarrow }i{\bf h}$, the same replacement
in the conventional $\sigma $-models would not lead to Eq. (\ref{a42}). This
reflects an essential symmetry difference between systems in a magnetic
field where the time reversal invariance is broken and the models with
direction that are time reversal invariant.

In contrast to average density of states for Hermitian disorder problems
which is always smooth, the joint probability density of complex
eigenenergies considered in the previous Sections is a non-trivial quantity.
The $\sigma $-model was derived to describe this quantity and it is expected
to be sensitive to localization-delocalization transitions in one- and
higher dimensional systems\cite{hatano}.

The the form of $0D$ version of the $\sigma $-model obtained above
demonstrates the equivalence between the directed disorder models in a
limited volume and ensembles of random weakly non-Hermitian or weakly
asymmetric real matrices that have been mapped onto the $0D$ $\sigma $-model
previously\cite{fyodorov}. Complex random non-Hermitian matrices appear in
study of dissipative quantum maps\cite{grobe,haake} whereas random real
asymmetric matrices have applications in neural network dynamics\cite
{som,doyon}. So, the $\sigma $-model can describe completely different
phenomena in an unified manner.

The supermatrix $\sigma $-model can serve as an useful calculational tool
for all these non-Hermiatian problems. Although the new term with the matrix 
$\Lambda _1$ in the $\sigma $-model, Eq. (\ref{a42}), makes the use of
previous parametrizations\cite{efetov} difficult, the new parametrization
suggested in the present paper allows to circumvent the difficulties and
obtain in a straightforward manner explicit results for the $0D$ case.
Weakly non-Hermitian random matrices can also be studied using more
traditional methods of orthogonal polynomials\cite{fyodorov1}. However,
study of weakly non-symmetric real matrices with this method seems be more
difficult and the density of complex eigenvalues, Eqs. (\ref{o28}-\ref{o30}%
), has been calculated for the first time. Besides, the $\sigma $-model
approach is not dependent on details of the model considered and can be
applied not only to Gaussian models. It can also be used to study the
directed models in one and higher dimensions where one can expect
localization-delocalization transitions.

Eqs. (\ref{o28}-\ref{o30}) demonstrate that at any finite disorder and
``imaginary vector-potential'' a finite portion of eigenvalues remain real
whereas this does not occur if the time reversal invariance is broken, Eq. (%
\ref{e25}). This phenomenon has manifested itself in numerical study of
different models. In Refs.\cite{sommers,leh} ensembles of random strongly
asymmetric matrices (symmetric and antisymmetric parts had the same order of
magnitude) were considered. It was found that the fraction of real
eigenvalues decayed as $N^{-1/2}$ for large matrix sizes $N$. Apparently,
this corresponds to the finite fraction of the real eigenvalues $P_r\left(
\epsilon ,y\right) $, Eq. (\ref{o28}), because in the ensemble of weakly
non-symmetric matrices involved the magnitude of the antisymmetric part of
the random matrices is $N^{1/2}$ times smaller than that of the symmetric
one.

A finite fraction of real eigenenergies was found in a numerical study of
the $2D$ model, Eq. (\ref{a2}), (without magnetic interactions) near the
center of the band\cite{hatano}. Although the $2D$ case was not considered
in the present paper and nothing can be said about a possibility of a
mixture of eigenstates with real and complex eigenvalues one can argue that,
may be, the parameters of the model of Ref.\cite{hatano} corresponded to the 
$0D$ case. This might easily happen because the localization length in
weakly disordered $2D$ systems is exponentially large and can exceed the
sample size, which would correspond to the $0D$ regime. If this is really
so, the results of the present study are in an agreement with the numerical
investigation.

The phenomenon that some finite portion of eigenvalues lies on a certain
line in the complex plane occurs also in other models with a randomness.
Recently, it was found that a finite fraction of all roots of random
self-inversive polynomials lies on the unit circle\cite{bogomolny}. At the
same time, if the polynomials are not self-inversive the density of complex
roots in smooth everywhere in the complex plane.

It is clear from the preceding discussion that the directed disorder models
deserve further investigation.

\section{Appendix}

\subsection{Non-Hermitian parametrization (unitary ensemble)}

Let us calculate for the unitary ensemble the Jacobian for the
parametrization given by Eqs.. (\ref{e7}-\ref{e11}) (it was suggested to
call it ``non-Hermitian parametrization''. As usual\cite{efetov0,efetov}, it
is convenient to consider the length $Str\left( dQ\right) ^2$. With Eq. (\ref
{e7}), it can be written as 
\begin{equation}
\label{ap1}STr\left( dQ\right) ^2=STr\left( \left( dQ_0\right) ^2+\left[
\delta T,Q_0\right] ^2+4\delta T\delta Q_0\right) 
\end{equation}
where $\delta T=\bar TdT$, $\delta Q_0=Q_0dQ_0$ and $\left[ .,.\right] $ is
the commutator.

It is easy to see from Eq. (\ref{e8}) that 
\begin{equation}
\label{ap2}\delta Q_0=\left( 
\begin{array}{cc}
0 & -\tau _3d 
\hat \varphi \\ \tau _3d\hat \varphi & 0 
\end{array}
\right) 
\end{equation}
and hence 
\begin{equation}
\label{ap3}\left\{ \delta Q_0,\Lambda _1\right\} =0 
\end{equation}
where $\left\{ .,.\right\} $ is the anticommutator.

Then, using the relation $\left[ \delta T,\Lambda _1\right] =0$ and Eq. (\ref
{ap3}) we obtain 
\begin{equation}
\label{ap3a}STr\left( \delta T\delta Q_0\right) =STr\left( \Lambda _1\delta
T\delta Q_0\Lambda _1\right) =-STr\left( \delta T\delta Q_0\right) =0 
\end{equation}
which shows that Jacobians is the product of Jacobians corresponding to $%
dQ_0 $ and $\delta T$. As concerns $dQ_0$, we have 
\begin{equation}
\label{ap5}STr\left( dQ_0\right) ^2=4\left( \left( d\varphi \right)
^2+\left( d\chi \right) ^2\right) 
\end{equation}
Writing Eq. (\ref{e9}) as 
\begin{equation}
\label{ap6}T=uT_0v 
\end{equation}
one obtains 
\begin{equation}
\label{ap7}\delta T=\bar v\bar T_0\delta uT_0v+\bar v\delta T_0v+\delta v 
\end{equation}
where, with Eq. (\ref{e10}, \ref{e11}) 
\begin{equation}
\label{ap8}\delta T_0=-\frac i2\left( 
\begin{array}{cc}
0 & d 
\hat \theta \\ d\hat \theta & 0 
\end{array}
\right) 
\end{equation}

$$
\delta u=\delta u_{\Vert }+\delta u_{\bot }, 
$$

$$
\delta u_{\Vert }=2\tau _3\left( \eta d\eta ^{*}-d\eta \eta ^{*}\right)
,\;\delta u_{\bot }=2\left( 
\begin{array}{cc}
0 & d\eta \\ 
-d\bar \eta & 0 
\end{array}
\right) 
$$
and similar equations can be written for $\delta v$.

Substituting Eq. (\ref{e10}, \ref{e11}, \ref{ap8}) into Eq. (\ref{ap7}) one
can represent the supermatrix $\delta T$ as 
\begin{equation}
\label{ap9}\delta T=\delta T^{\Vert }+\delta T^{\bot }, 
\end{equation}
\begin{equation}
\label{ap10}\delta T^{\Vert }=2\cos \frac{\theta -i\theta _1}2\left( 
\begin{array}{cc}
0 & d\eta \\ 
-d\bar \eta & 0 
\end{array}
\right) +2\left( 
\begin{array}{cc}
0 & d\kappa \\ 
-d\bar \kappa & 0 
\end{array}
\right) 
\end{equation}

$$
+2\tau _3\left( \eta d\eta ^{*}-d\eta \eta ^{*}+\kappa d\kappa ^{*}-d\kappa
\kappa ^{*}\right) +4\tau _3\cos \frac{\theta -i\theta _1}2\left( \kappa
^{*}d\eta -d\eta ^{*}\kappa \right) , 
$$
\begin{equation}
\label{ap11}\delta T^{\bot }=i\Lambda _1(2\sin \frac{\theta -i\theta _1}2%
\left( 
\begin{array}{cc}
0 & d\eta \\ 
d\bar \eta & 0 
\end{array}
\right) -\left( d\theta -id\theta _1\right) \left( 
\begin{array}{cc}
0 & \kappa \\ 
\bar \kappa & 0 
\end{array}
\right) 
\end{equation}

$$
-\frac 12\left( 
\begin{array}{cc}
d\theta \left( 1-4\kappa \kappa ^{*}\right) & 0 \\ 
0 & id\theta _1\left( 1+4\kappa \kappa ^{*}\right) 
\end{array}
\right) +4\sin \frac{\theta -i\theta _1}2\left( \kappa ^{*}d\eta +d\eta
^{*}\kappa \right) ) 
$$
In Eqs. (\ref{ap9}-\ref{ap11}) $\delta T^{\Vert }$ commutes with $\Lambda $, 
$\delta T^{\bot }$ anticommutes with $\Lambda $. The second line in Eq. (\ref
{ap10}) does not contribute to $\left[ \delta T,Q_0\right] $ in Eq. (\ref
{ap1}). In Eqs. (\ref{ap10}, \ref{ap11}), one can change the variables 
\begin{equation}
\label{ap12}d\theta \left( 1-4\kappa \kappa ^{*}\right) \rightarrow d\theta
,\;d\theta _1\left( 1+4\kappa \kappa ^{*}\right) \rightarrow d\theta _1 
\end{equation}
and make the shifts 
\begin{equation}
\label{ap13}\frac 12\left( 
\begin{array}{cc}
d\theta & 0 \\ 
0 & id\theta _1 
\end{array}
\right) \rightarrow \frac 12\left( 
\begin{array}{cc}
d\theta & 0 \\ 
0 & id\theta _1 
\end{array}
\right) +4\left( \kappa ^{*}d\eta +d\eta ^{*}\kappa \right) \sin \frac{%
\theta -i\theta _1}2 
\end{equation}
\begin{equation}
\label{ap14}2\sin \frac{\theta -i\theta _1}2d\eta \rightarrow 2\sin \frac{%
\theta -i\theta _1}2d\eta +\left( d\theta -id\theta _1\right) \kappa 
\end{equation}
\begin{equation}
\label{ap15}d\kappa \rightarrow d\kappa -\cos \frac{\theta -i\theta _1}2%
d\eta 
\end{equation}

The transformations, Eqs. (\ref{ap12}-\ref{ap15}) do not change the Jacobian
and $\delta T^{\Vert }$ and $\delta T^{\bot }$ take a more simple form 
\begin{equation}
\label{ap16}\delta T^{\Vert }={\bf 1}\left( i\tau _3\left( 
\begin{array}{cc}
dc & 0 \\ 
0 & dc 
\end{array}
\right) +2\left( 
\begin{array}{cc}
0 & d\kappa \\ 
-d\bar \kappa & 0 
\end{array}
\right) \right) 
\end{equation}
\begin{equation}
\label{ap17}\delta T^{\bot }=i\Lambda _1\left( -\frac 12\left( 
\begin{array}{cc}
d\theta & 0 \\ 
0 & id\theta _1 
\end{array}
\right) +2\sin \frac{\theta -i\theta _1}2\left( 
\begin{array}{cc}
0 & d\eta \\ 
d\bar \eta & 0 
\end{array}
\right) \right) 
\end{equation}
where $i\tau _3dc$ is the second line of Eq. (\ref{ap10}) and ${\bf 1}$ is
the unit $8\times 8$ matrix.

Further computation is already simple. Changing once more 
\begin{equation}
\label{ap18}\sin \frac{\theta -\theta _1}2d\eta \rightarrow d\eta 
\end{equation}
one obtains a contribution to the Jacobian proportional to $J_\theta $, Eq. (%
\ref{e16a}). Writing in the new variables the second term in Eq. (\ref{ap1})
we have 
\begin{equation}
\label{ap19}STr\left[ \delta T,Q_0\right] ^2=4\left( \left( d\theta \right)
^2\cos ^2\varphi +\left( d\theta _1\right) ^2\cosh ^2\chi \right) 
\end{equation}

$$
+128\left( \cos ^2\frac{\varphi +i\chi }2d\eta d\eta ^{*}+\sin ^2\frac{%
\varphi -i\chi }2d\kappa d\kappa ^{*}\right) 
$$
Eqs. (\ref{ap5}, \ref{ap19}) lead to the elementary volume $\left[ dQ\right] 
$, Eq. (\ref{e15}).

\subsection{Non-symmetric parametrization (orthogonal ensemble)}

To calculate the Jacobian of the parametrization, Eqs. (\ref{o1}-\ref{o7}),
for the orthogonal ensemble we can use the results obtained for the unitary
ensemble because Eq. (\ref{o1}) contains the same supermatrices $Q_0$ and $T$
as previously. However, the presence of the supermatrix $Y$ makes the
computation quite lengthy. The length $STr\left( dQ\right) ^2$ is written as
in the preceding Subsection 
\begin{equation}
\label{ap20}STr\left( dQ\right) ^2=STr\left( \left( dQ_0\right) ^2+\left[
\delta Z,Q_0\right] ^2+4\delta Z\delta Q_0\right) 
\end{equation}
where $\delta Z=\bar ZdZ$ can be written as 
\begin{equation}
\label{ap21}\delta Z=\bar S\bar R\left( \bar Y_0\delta TY_0+\delta Y_0+dR%
\bar R+RdS\bar S\bar R\right) RS 
\end{equation}
The last term in Eq. (\ref{ap20}) is equal to zero (see Eq. (\ref{ap3a})).
As concerns the supermatrix $\delta T$, it can be written after the
replacements, Eqs. (\ref{ap12}-\ref{ap15}), in the form of Eqs. (\ref{ap16}, 
\ref{ap17}). So, one has to calculate the other differentials entering Eq. (%
\ref{ap21}). Using Eq. (\ref{o7}) one can rewrite $dS\bar S$ to the form 
\begin{equation}
\label{ap22}dS\bar S=\left( dS\bar S\right) _{\Vert }+\left( dS\bar S\right)
_{\bot }, 
\end{equation}
$$
\left( dS\bar S\right) _{\Vert }=2\tau _3{\bf 1}\left( d\sigma \sigma
^{*}-\sigma d\sigma ^{*}\right) ,\;\left( dS\bar S\right) _{\bot }=2i\Lambda
_1d\hat \sigma 
$$
Taking the supermatrix $R$ from Eq. (\ref{o6}) one can derive 
\begin{equation}
\label{ap23}RdS\bar S\bar R=R\left( dS\bar S\right) _{\bot }\bar R+\left( dS%
\bar S\right) _{\Vert }, 
\end{equation}
\begin{equation}
\label{ap23a}R\left( dS\bar S\right) _{\bot }\bar R=2i\Lambda _1d\hat \sigma
+4i\Lambda _1\left( d\sigma \rho ^{*}+\rho d\sigma ^{*}\right) 
\end{equation}
and 
\begin{equation}
\label{ap24}dR\bar R={\bf 1}\left( 2\left( 
\begin{array}{cc}
0 & d\rho \\ 
-d\bar \rho & 0 
\end{array}
\right) +2\tau _3\left( d\rho \rho ^{*}-\rho d\rho ^{*}\right) \right) 
\end{equation}
Now we have to calculate $\delta Y_0$. Using Eqs. (\ref{o3}-\ref{o5}) one
can represent this differential in the form 
\begin{equation}
\label{ap25}\delta Y_0=\delta Y_1+\delta Y_2+\bar Y_1\bar Y_2\delta
Y_3Y_2Y_1 
\end{equation}
Calculating the matrices $\delta Y_1,\delta Y_2$, and $\delta Y_3$ we
rewrite $\delta Y_0$ as follows%
$$
\delta Y_0={\bf 1}\left( \frac i2\left( 
\begin{array}{cc}
d\beta \bar w\tau _3w & 0 \\ 
0 & d\beta _1\tau _3\cosh \theta _2 
\end{array}
\right) -\frac i2d\mu \left( 
\begin{array}{cc}
\tau _2 & 0 \\ 
0 & 0 
\end{array}
\right) \right) 
$$
\begin{equation}
\label{ap26}+\Lambda _1\left( \frac i2\left( 
\begin{array}{cc}
0 & 0 \\ 
0 & \tau _2d\beta _2 
\end{array}
\right) +\frac 12d\theta _2\left( 
\begin{array}{cc}
0 & 0 \\ 
0 & \tau _1 
\end{array}
\right) \right) 
\end{equation}
where 
$$
\tau _2=\left( 
\begin{array}{cc}
0 & -i \\ 
i & 0 
\end{array}
\right) ,\;\bar w\tau _3w=\left( 
\begin{array}{cc}
\cos \mu & -\sin \mu \\ 
-\sin \mu & -\cos \mu 
\end{array}
\right) 
$$
Making the replacement%
$$
d\kappa \rightarrow d\kappa \exp \frac{i\left( \beta -\beta _1\right) }2%
,\;d\kappa ^{*}\rightarrow d\kappa ^{*}\exp \frac{i\left( \beta -\beta
_1\right) }2 
$$
and the same for $d\eta $ and $d\eta ^{*}$ one can derive 
\begin{equation}
\label{ap27}\bar Y_0\delta TY_0=2{\bf 1}\left( \cosh \frac{\theta _2}2\left( 
\begin{array}{cc}
0 & d\kappa ^{\prime } \\ 
-d\kappa ^{\prime } & 0 
\end{array}
\right) +i\sinh \frac{\theta _2}2\left( 
\begin{array}{cc}
0 & d\eta ^{\prime }\tau _1 \\ 
-\tau _1d\bar \eta ^{\prime } & 0 
\end{array}
\right) \right) 
\end{equation}

$$
+2i\Lambda _1\left( \cos \frac{\theta _2}2\left( 
\begin{array}{cc}
0 & d\eta ^{\prime } \\ 
d\bar \eta ^{\prime } & 0 
\end{array}
\right) -i\sinh \frac{\theta _2}2\left( 
\begin{array}{cc}
0 & d\kappa ^{\prime }\tau _1 \\ 
\tau _1d\bar \kappa ^{\prime } & 0 
\end{array}
\right) -\frac i2d\hat \theta \right) 
$$
where $d\eta ^{\prime }=\bar wd\eta $, $d\bar \eta ^{\prime }=d\bar \eta w$
and the same for $d\kappa ^{\prime }$ and $d\bar \kappa $. The contribution
from $i\tau _2dc$, Eq. (\ref{ap16}), is not written because it can be
removed by a proper shift of $d\hat \beta $ and $d\mu $.

Substituting Eqs. (\ref{ap22}-\ref{ap24}) into Eqs. (\ref{ap21}, \ref{ap20})
we see that second terms of Eq. (\ref{ap23}, \ref{ap24}) do not contribute.
After making the replacement in $\delta T$, Eq. (\ref{ap18}), and shifting 
\begin{equation}
\label{ap28}d\sigma =d\sigma _1-\left( \cosh \frac{\theta _2}2\cos \frac \mu 
2d\eta +i\sinh \frac{\theta _2}2\sin \frac \mu 2d\kappa ^{*}\right) 
\end{equation}

$$
d\rho =d\rho _1-\left( \cosh \frac{\theta _2}2\cos \frac \mu 2d\kappa
-i\sinh \frac{\theta _2}2\sin \frac \mu 2d\eta ^{*}\right)  
$$
it is convenient to introduce the matrix differentials 
\begin{equation}
\label{ap29}d\sigma =\left( 
\begin{array}{cc}
d\sigma _1 & d\sigma _2 \\ 
-d\sigma _2^{*} & -d\sigma _1^{*}
\end{array}
\right) ,\;d\rho =\left( 
\begin{array}{cc}
d\rho _1 & d\rho _2 \\ 
-d\rho _2^{*} & -d\rho _1^{*}
\end{array}
\right) 
\end{equation}
where 
\begin{equation}
\label{ap30}d\sigma _2=-\cosh \frac{\theta _2}2\sin \frac \mu 2d\eta
^{*}-i\sinh \frac{\theta _2}2\cos \frac \mu 2d\kappa 
\end{equation}
$$
d\sigma _2^{*}=\cosh \frac{\theta _2}2\sin \frac \mu 2d\eta -i\sinh \frac{%
\theta _2}2\cos \frac \mu 2d\kappa ^{*} 
$$
$$
d\rho _2=-\cosh \frac{\theta _2}2\sin \frac \mu 2d\kappa ^{*}+i\sinh \frac{%
\theta _2}2\cos \frac \mu 2d\eta  
$$
$$
d\rho _2^{*}=\cosh \frac{\theta _2}2\sin \frac \mu 2d\kappa +i\sinh \frac{%
\theta _2}2\cos \frac \mu 2d\eta ^{*} 
$$
The Jacobian $\tilde J_\mu $ of the transformation, Eqs. (\ref{a30}), equals 
\begin{equation}
\label{ap31}\tilde J_\mu =\frac 4{\left( \cosh \theta _2-\cos \mu \right) ^2}
\end{equation}
Then Eq. (\ref{ap21}) can be written as 
\begin{equation}
\label{ap32}\delta Z=\bar S\bar R\delta URS,
\end{equation}
\begin{equation}
\label{ap33}\delta U=\delta Y_0+i\Lambda _1\left( 2d\hat \sigma -\frac i2d%
\hat \theta \right) +2kd\hat \rho {\bf 1}
\end{equation}
where the matrices $d\sigma $ and $d\rho $ entering $d\hat \sigma $ and $d%
\hat \rho $, Eq. (\ref{o7}) have the structure Eq. (\ref{ap29}) and%
$$
k=\left( 
\begin{array}{cc}
1 & 0 \\ 
0 & -1
\end{array}
\right)  
$$
. One can calculate $\delta Z$, Eq. (\ref{ap32}), calculating first $\bar R%
\delta UR$ and then $\delta Z$. The corresponding manipulations are still
quite lengthy. One should again make different replacements that do not
change the Jacobian. Alternatively, one might write the final result using
general symmetry properties of $\delta Z$. Finally one obtains 
\begin{equation}
\label{ap34}\delta Z=\delta Y_0^{\prime }+i\Lambda _1\left( 2d\hat \sigma -%
\frac 12d\hat \theta \right) +2kd\hat \rho {\bf 1}
\end{equation}
The supermatrix $\delta Y_0^{\prime }$ entering Eq. (\ref{ap34}) equals 
\begin{equation}
\label{ap35}\delta Y_0^{\prime }{\bf =-1}\frac i2\left( d\beta \sin \mu
\left( 
\begin{array}{cc}
\tau _1 & 0 \\ 
0 & 0
\end{array}
\right) \right) +d\mu \left( 
\begin{array}{cc}
\tau _2 & 0 \\ 
0 & 0
\end{array}
\right) 
\end{equation}

$$
+\Lambda _1\frac 12\left( -\sinh \theta _2d\beta _1\left( 
\begin{array}{cc}
0 & 0 \\ 
0 & \tau _2
\end{array}
\right) +d\theta _2\left( 
\begin{array}{cc}
0 & 0 \\ 
0 & \tau _1
\end{array}
\right) \right)  
$$
Using Eq. (\ref{ap34}) we can calculate $STr\left[ \delta Z,Q_0\right] ^2$.
The contribution of the anticommuting part $\delta Z_{\bot }$ decouples from
that of the commuting one $\delta Z_{\Vert }$ and one obtains
\begin{equation}
\label{ap36}STr\left[ \delta Z_{\bot },Q_0\right] ^2=64[d\sigma _1d\sigma
_1^{*}\left( 1+\cos \left( \varphi +i\chi \right) \right) +d\sigma _2d\sigma
_2^{*}\left( 1+\cos \left( \varphi -i\chi \right) \right) 
\end{equation}
$$
+d\rho _1d\rho _1^{*}\left( 1-\cos \left( \varphi -i\chi \right) \right)
+d\rho _2d\rho _2^{*}\left( 1-\cos \left( \varphi +i\chi \right) \right)  
$$
The Jacobian $J_{\varphi \chi }$ corresponding to the length, Eq. (\ref{ap36}%
) equals
\begin{equation}
\label{ap37}J_{\varphi \chi }=\frac 1{2^{24}}\frac 1{\left( \sin ^2\varphi
+\sinh ^2\chi \right) ^2}
\end{equation}
The commuting part $\delta Z_{\Vert }$ contributes to the elementary length
as
\begin{equation}
\label{ap38}STr\left[ \delta Z_{\Vert },Q_0\right] ^2=4[\left( \left( d\mu
\right) ^2+\left( d\beta \right) ^2\sin ^2\mu \right) \sin ^2\varphi 
\end{equation}

$$
+\left( d\theta \right) ^2\cos ^2\varphi +\left( d\theta _1\right) ^2\cosh
^2\chi +\left( d\theta _2\right) ^2+\left( d\beta _1\right) ^2\sinh ^2\theta
_2] 
$$

Combining the contribution coming to the Jacobian from Eqs. (\ref{ap5}, \ref
{ap38}) with those written in Eqs. (\ref{ap31}, \ref{ap37}) and recalling
that the replacement, Eq. (\ref{ap18}), results in an additional multiplier
proportional to $J_\theta $ one obtains finally the elementary volume $%
\left[ dQ\right] $, Eqs. (\ref{o9}-\ref{o12}, \ref{e15}-\ref{e16a}).

\end{document}